\documentclass[12pt,preprint,letter]{aastex}

\slugcomment{}

\DeclareGraphicsExtensions{.eps}
\newcommand{\be}{\begin{equation}}
\newcommand{\ee}{\end{equation}}
\newcommand{\ba}{\begin{eqnarray}}
\newcommand{\ea}{\end{eqnarray}}
\newcommand{\bc}{\begin{center}}
\newcommand{\ec}{\end{center}}

\def\Lsun{L$_\odot$}
\def\Msun{M$_\odot$}

\begin{document}

\title{Theoretical modelling of
the diffuse emission of $\gamma$-rays from extreme regions of star
formation:  The case of Arp 220}

\author{Diego F. Torres}

\affil{\rm Lawrence Livermore National Laboratory \\ 7000 East
Ave. L$-$413, Livermore, CA 94550. E-mail:
dtorres@igpp.ucllnl.org}

\begin{abstract}

\begin{center}
{\bf Astrophys.J.617:966-986, 2004}\end{center}

\mbox \\ Our current understanding of ultraluminous infrared
galaxies suggest that they are recent galaxy mergers in which much
of the gas in the former spiral disks, particularly that located at
distances less than 5 kpc from each of the pre-merger nuclei, has
fallen into a common center, triggering a huge starburst phenomenon.
This large nuclear concentration of molecular gas has been detected
by many groups, and estimates of molecular mass and density have
been made. Not surprisingly, these estimates were found to be orders
of magnitude larger than the corresponding values found in our
Galaxy. In this paper, a self-consistent model of the high energy
emission of the super-starburst galaxy Arp 220 is presented. The
model also provides an estimate of the radio emission from each of
the components of the central region of the galaxy (western and
eastern extreme starbursts, and molecular disk). The predicted radio
spectrum is found as a result of the synchrotron and free-free
emission, and absorption, of the primary and secondary steady
population of electrons and positrons. The latter is output of
charged pion decay and knock-on leptonic production, subject to a
full set of losses in the interstellar medium. The resulting radio
spectrum is in agreement with sub-arcsec radio observations, what
allows to estimate the magnetic field. In addition, the FIR emission
is modeled with dust emissivity, and the computed FIR photon density
is used as a target for inverse Compton process as well as to give
account of losses in the $\gamma$-ray scape. Bremsstrahlung emission
and neutral pion decay are also computed, and the $\gamma$-ray
spectrum is finally predicted. Future possible observations with
GLAST, and the ground based Cherenkov telescopes are discussed.

\end{abstract}


\section{Introduction }

In a recent letter (Torres et al. 2004),  it was shown that some
luminous and ultra-luminous infrared galaxies (LIRGs and ULIRGs) are
plausible sources for GLAST and the next generation of Cherenkov
telescopes (HESS, MAGIC, VERITAS). In order to show that, the
$\gamma$-ray flux output of neutral pion decay, under a set of
reasonable and commonly used --albeit numerous-- simplifications,
was computed. An obvious caveat of this earlier approach is that it
was not possible to predict an spectrum of the ULIRGs-emitted
high-energy radiation, but rather only integrated fluxes. Also,
correlation at lower frequencies was not pursued. Here, a detailed,
self-consistent model of the radio, IR, and $\gamma$-ray emission
from Arp 220, the nearest ULIRG, minimizing as much as possible
--based on current multiwavelength observations-- any freedom in
parameter selection, is presented.

To that end, a set of numerical codes that allow the computation of
multiwavelength spectra from regions of star formation, molecular
clouds, and other environments, was developed. Being this the first
application of such program --whose validation was run against
previously published results-- some of the details of what it
implements are discussed in a technical Appendix. The code set,
dubbed ${\cal Q}$-{\sc diffuse}, solves the diffusion-loss equation
for electrons and protons, and finds the steady state distribution
for these particles subject to a complete set of losses in the
interstellar medium (ISM). It computes secondaries from hadronic
interactions (neutral and charged pions) and Coulomb processes
(electrons), and gives account of the radiation or decay products
that these particles produce. Secondary particles (photons, muons,
neutrinos, electrons, and positrons) that are in turn produced by
pion decay  are calculated too, using a new set of paramaterizations
of the differential cross sections, developed recently by Blattnig
et al. (2000). These parameterizations are discussed here in some
detail as well. Additional pieces of the code compute the dust
emissivity, and the IR-FIR photon density, which is used both as
target for inverse Compton scattering and to model the radiation at
lower frequencies. Finally, opacities to $\gamma \gamma$ and $\gamma
Z$ processes are computed, as well as absorbed $\gamma$-ray fluxes,
using the radiation transport equation.

Previous studies of diffuse high energy emission, and of electron
and positron production, with different levels of detail and aims,
go back to the early years of $\gamma$-ray astronomy. A summary of
these first efforts can be found in the review paper by Fazio (1967)
and in the book by Ginzburg and Syrovatskii (1968). See also the
pioneering works by Ramaty \& Lingenfelter (1968), Maraschi et al.
(1968), and Stecker (1977), among many others. Secondary particle
computations have a similarly long, and obviously related history
see, e.g., Stecker (1969; 1973), Orth and Buffington (1976), and
others quoted below. More recent efforts, related mainly to the
modelling of supernova remnants and the Galactic center, include
those of Schlickeiser (1982), see also his book and references
quoted therein (Schlickeiser 2002), Aharonian et al. (1994), Drury
et al. (1994), Atoyan et al. (1995), Aharonian \& Atoyan (1996),
Moskalenko \& Strong (1998), Strong \& Moskalenko (1998), Markoff et
al. (1999), and Fatuzzo \& Melia (2003); although making here a
comprehensive list is not intended. Here, the general ideas used by
Paglione et al. (1996) and Blom et al. (1999), when modelling nearby
starbursts galaxies, are followed. These, in turn, closely track
Brown \& Marscher's (1977) and Marscher \& Brown's (1978), regarding
their studies of close molecular clouds. The current implementation
seems to introduce some further improvements. Apart from using
different parameterizations for pion cross sections, which were
argued to better agree with experiments, as mentioned above, the
code set uses the full inverse Compton Klein-Nishina cross section,
computes secondaries without resorting to parameterizations which
are valid only for Earth-like cosmic ray (CR) intensities, fixes the
photon target for Compton scattering starting from modelling of the
observations in the FIR, and considers opacities to $\gamma$-ray
scape.

The rest of this paper is organized as follows. In the next Section,
LIRGs and ULIRGs as $\gamma$-ray sources are discussed. Section 3 is
an account of Arp 220 phenomenology. The description of the dust
emission model and the supernova explosion rates that were
implemented are discussed there as well. Section 4 is a discussion
of the solution to the diffusion-loss equation in a general case.
Section 5 shows how emissivities of secondary particles were
computed. Section 6 discusses the steady distribution of particles
in the different components of Arp 220, together with the resulting
radio and $\gamma$-ray spectrum. Some concluding remarks are given
at the end.

\section{LIRGs \& ULIRGs as $\gamma$-ray sources}

ULIRGs are recent galaxy mergers in which much of the gas in the
former spiral disks, particularly that located at distances less
than $\sim 5$ kpc from each of the pre-merger nuclei, has fallen
into a common center, triggering a huge starburst phenomenon (see
Sanders \& Mirabel 1996 for a review). The size of the inner regions
of ULIRGs, where most of the gas is found, can be as small as a few
hundreds parsecs; there, an extreme molecular environment is found.

This large nuclear concentration of molecular gas has been detected
in the millimeter lines of CO by many groups. Using Milky Way
molecular clouds to calibrate the conversion factor between CO
luminosity and gas mass soon led to the paradox that most, if not
all, of the dynamical mass was gas (e.g., for Arp 220, see Scoville
et al. 1991). In some extreme cases, the derived gas mass exceeded
the dynamical mass estimation, which unambiguously showed caveats in
any of the assumptions. However, Downes et al. (1993) showed that in
the central regions of ULIRGs, much of the CO luminosity comes from
an intercloud medium that fills the whole volume, rather than from
clouds bound by self gravity. Hence, the CO luminosity of ULIRGs
traces the geometric mean of the gas and the dynamical mass, rather
than just the gas. The Milky Way conversion factor, being relevant
for an ensemble of giant molecular clouds (GMCs) in an ordinary
spiral galaxy, seems to overestimate the gas mass of ULIRGs. Solomon
et al. (1997), Downes \& Solomon (1998), Bryant \& Scoville (1999),
and Yao et al. (2003) have argued for that in the case of ULIRGs,
conversion factors between gas mass and CO luminosities can be
$\sim$5 times smaller than for the Milky Way. Even with such
corrections, the amount of molecular gas in ULIRGs is huge,
typically reaching 10$^{10}$ M$_\odot$.

The existence of large masses of dense interstellar gas suggests
that all LIRGs may have $\gamma$-ray luminosities orders of
magnitude greater than normal galaxies. This assumption was explored
by Torres et al. (2004), who found that the expectation of LIRGs to
shine at $\gamma$-rays is not automatically granted. It is not only
the amount of gas (actually, the amount of gas divided by the
distance to its location) what yields to detectability at high
energies, but rather it is the amount of gas that is found at high
density, and thus that it is prone to form stars and be subject to
significant enhancements of cosmic rays. Using the HCN survey
recently released by Gao \& Solomon (2004a,b), Torres et al. noted
that there are a group of 7 LIRGs (out of 31 in that sample) that,
being gas-rich (i.e., CO-luminous) but having normal star formation
efficiency $L_{\rm IR}/L_{\rm CO}$ (e.g., $L_{\rm HCN}/L_{\rm CO} <
0.06$), are not expected to be detected in $\gamma$-rays (at least
under the simple modelling explored by these authors). Some examples
are NGC 1144, Mrk 1027, NGC 6701, and Arp 55. These galaxies are
using the huge molecular mass they have in creating stars at a
normal star formation rate (SFR). Cosmic ray enhancements are, most
likely, not high enough to lead to detection, given the distance to
these objects.

Then, even when they may appear far from Earth to be detected at
high energies, perhaps it is the extreme environment of
star-bursting ULIRGs the most appealing to study. One such galaxy
stands alone among all others: Arp 220 (RA$_{\rm J2000}$, DEC$_{\rm
J2000}$=15 34 57.24, +23 30 11.2).
%
%
Although LIRGs are the dominant population of extragalactic objects
in the local ($z<0.3$) universe at bolometric luminosities above $L
> 10^{11}$ L$_\odot$, they are still relatively rare (Sanders \&
Mirabel 1996). The luminosity function of LIRGs suggest that there
should be only one object with $L_{\rm FIR}>10^{12}$ L$_\odot$ out
to a redshift of 0.033. Indeed, Arp 220 ($z=0.018$) is the only
ULIRG in the 100 Mpc sphere. As such, Arp 220 is probably the best
studied ULIRG.

\section{Arp 220}

Arp 220's center has two radio-continuum and two IR sources,
separated by $\sim 1$ arcsec (e.g., Scoville et al. 1997, Downes et
al. 1998, Soifer et al. 1999, Wiedner et al. 2002). The two radio
sources are extended and nonthermal (e.g., Sopp \& Alexander 1991;
Condon et al. 1991; Baan \& Haschick 1995), and likely produced by
supernovae in the most active star-forming regions. CO line, cm,
mm-, and sub-mm continuum (e.g., Downes \& Solomon 1998) as well as
recent HCN line observations (e.g., Gao \& Solomon 2004a,b) are all
consistent with these two sources being sites of extreme star
formation and having very high molecular densities. Arp 220 is also
an OH megamaser galaxy, as first discover by Baan et al. (1982). The
1.6 GHz continuum emission of Arp 220 has a double component
structure too, with the two components being separated by about 1
arcsec and located at the same positions as the 1.4 GHz, the 4.8
GHz, and the 1.3 mm emission (see, e.g., Rovilos et al. 2002, 2003).
In the eastern nucleus, the position of the maser coincide with that
of the continuum. In the western one, the OH maser emission arises
from regions north and south from the continuum (Rovilos et al.
2002, 2003).

Different characteristics of the two extreme starbursts and the
molecular disk, some of which are used as input in our modelling,
are given in Tables 1 and 2, as derived by Downes and Solomon
(1998). Other authors, particularly those reporting results with
sub-arcsec angular resolution (e.g., Soifer et al. 1999, Wiedner et
al. 2002), while confirming the general features of the modelling of
the central region proposed by Downes and Solomon, may present
differences in the details. For instance, the densities quoted by
Weidner et al. (2002) are slightly larger than those used here.
Sakamoto et al. (1999) have proposed, also based on CO observations
with sub-arcsec resolution, that the western and eastern nuclei are
not spherically symmetric but are counter-rotating, $\sim 100$ pc
disks, with $\sim 10^9$ M$_\odot$ masses (see their figure 5). This
model seems to have some support in VLBI observations of OH masers
(Rovilos et al. 2003). Regarding the $\gamma$-ray emission from Arp
220, such changes in geometry will not yield any significant change
in the results, although would probably also imply higher densities
that those consider here. To fix the scenario on the conservative
side, Downes and Solomon's (1998) results are adopted, and for
consistency, their assumed value of Arp 220 luminosity distance
(72.3 Mpc) is also used. Modifications to the cosmological model
would produce an order 1\% percent change in the results.

\begin{figure*}[t]
\centering
\includegraphics[width=.4\textwidth]{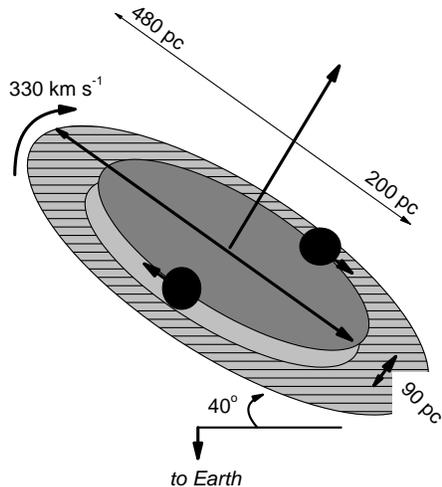}\hspace{0.5cm}
\caption{Geometry and different components in the model of Arp 220.
Two central spherical {\it nuclei} are extreme regions of star
formation, and co-rotate with the molecular disk.  } \label{geom}
\end{figure*}

The assumed geometry of the central region of Arp 220 is sketched in
Figure \ref{geom}, not to scale. The CO disk is inclined  40$^\circ$
from face-on, Arp~220-west (one of the extreme starbursts) is
assumed spherical, with a radius of 68\,pc. Similarly, Arp~220-east
has a radius of 110\,pc. The disk thickness is 90\,pc. The
rotational curve of the CO disk indicates a dynamical mass of at
least $12 \times 10^9$\,\Msun\ interior to the outer disk radius, of
480\,pc, which corresponds to the central bulge mass of a large
spiral like the Milky Way. The gas mass in each of the two extreme
starburst nuclei is at least $6\times 10^8$\,\Msun . Their
individual FIR luminosities are $\sim 3\times 10^{11}$\,\Lsun .
About half of the Arp~220 FIR luminosity comes from the molecular
disk. The masses of the two extreme starbursts are negligible in
comparison with the mass that controls the motion of the molecular
disk. Furthermore, the two nuclei of Arp~220 have radial velocities
indicating that they take part in the general disk rotation, i.e.,
that they share the general rotation in the potential of the old
bulge, and are dominated by the disk gravity, not their self. There
is no observational evidence --radio, infrared, or optical-- that
they contain old stars, so that the estimated mass in new stars
could just be the total mass minus the gas mass (Downes \& Solomon
1998). The gas density quoted in Table 1 and 2 corresponds only to
estimates of molecular hydrogen, thus the total density ought to be
larger.  The contribution of atomic hydrogen is to be considered
subdominant, as it is in the inner disk of the Milky Way (see e.g.
Mirabel \& Sanders 1988; 1989). The total nuclei density is derived
from the H$_2$ number density estimation, taking into account
heavier and lighter species. Also, it is important to note that the
models in the paper by Downes \& Solomon (1998) are for distributed
gas, but there is denser gas in the star forming cores, giving rise
to HCN and CS lines. Most of the CO comes from the distributed
medium, so that total masses have to be corrected upwards (e.g. Gao
\& Solomon 2004a,b). Equally, the density might be higher that the
estimate used here, perhaps especially in the disk. Thus, from the
point of view of target mass, our estimates of, for instance,
neutral pion decay $\gamma$-rays or charged pion decay electrons,
could be regarded as a conservative estimation.



\begin{deluxetable}{lll}
\tablewidth{0.8\textwidth}
\tablecaption{Some properties of Arp 220's extreme starbursts.} \tablehead{
\colhead{Property} & \colhead{West } & \colhead{East } }
\startdata

Geometry & sphere & sphere \nl

Radius [pc] & 68 & 110 \nl


Average gas density (H$_2$) [cm$^{-3}$] & $1.8 \times 10^4$ & $8.0
\times 10^3$ \nl



Luminosity (FIR) [L$_\odot$] & $0.3 \times 10^{12}$ & $0.2 \times
10^{12}$ \nl

\enddata
\end{deluxetable}

\begin{deluxetable}{ll}
\tablewidth{0.8\textwidth}
\tablecaption{Some properties of Arp 220's disk.}

\tablehead{ \colhead{Property} & \colhead{Value }

} \startdata

Geometry & cylinder  \nl

Thickness [pc] & 90 \nl


Outer radius [pc] & 480 \nl

Inclination from face-on& 40$^{\rm o}$  \nl




Average gas density within the outer radius (H$_2$) [cm$^{-3}$] &
$1.2 \times 10^3$ \nl

Luminosity (FIR) [L$_\odot$] & $0.7 \times 10^{12}$ \nl

\enddata
\end{deluxetable}

\begin{deluxetable}{lll}
\tablewidth{0.75\textwidth}
\tablecaption{Main symbols used in the paper, meaning, and units.}
\tablehead{ \colhead{Symbol} & \colhead{Meaning} & \colhead{Unit}
} \startdata

$b(E)$  & rates of energy loss  & GeV s$^{-1}$ \nl

$ \tau(E)$ & confinement timescales  & s \nl

$Q(E)$  & emissivities & particles GeV$^{-1}$ s$^{-1}$ cm$^{-3}$
\nl

$N(E)$ & distributions & particles GeV$^{-1}$ cm$^{-3}$ \nl

$J(E)$  & intensities & particles GeV$^{-1}$ cm$^{-2}$ s$^{-1}$
sr$^{-1}$\nl

$F(E)$ & differential fluxes & particles GeV$^{-1}$ cm$^{-2}$
s$^{-1}$ \nl

$F(E>\bar E)$ & integral fluxes above $\bar E$  & particles
cm$^{-2}$ s$^{-1}$ \nl

\enddata
\label{units}
\end{deluxetable}

Additional evidence supporting the dominance of star forming
processes in Arp 220, as compared with what would be the influence
of an active but hidden black hole, come from the hard X-ray
band/soft $\gamma$-ray bands. Dermer et al. (1997) have reported
OSSE observations of Arp 220, finding a $2\sigma$ upper limit in the
50-200 keV range (see below). Previous hard X-ray limits on Arp 220,
by HEAO-1 and Ginga (Rieke 1988) also ruled out a bright hard X-ray
source ($> {\rm few} \, \times  10^{-11}$ erg cm$^{-2}$ s$^{-1}$).
Iwasawa et al. (2001) reported observations with Beppo-Sax, which
detected X-ray emission up to 10 keV but imposed only an upper limit
at higher frequencies. It is also worth noticing that there is no
strong Fe K line detection from Arp 220, although a tentative {\em
detection} of an emission line at 6.5 keV, at the 2$\sigma$-level,
has been made (Clements et al. 2002).


Starburst phenomena were used by Shioya, Trentham \& Tanigushi
(2001) and Iwasawa et al. (2001) to explain the X-ray properties of
Arp 220, although the existence of a heavily obscured AGN is not yet
ruled out. Chandra results (Clements et al. 2002) show that the
nuclear X-ray emission in Arp 220 is confined to a sub-kiloparsec
scale region, in contrast to other starburst galaxies. Its spectrum
indicates that X-rays are more likely produced by one or more low
luminosity, heavily obscured, low mass AGN, or by several high
luminosity X-ray binaries, or ultra luminous X-ray sources, rather
than by supernovae. Therefore the co-existence of a subdominant AGN
with a dominant starburst is still plausible. Of course, even when a
weak AGN would contribute now only with $\sim 1\%$ to the bolometric
luminosity, in the dense nuclear region of Arp 220, the black hole
is bound to grow and increase in luminosity as the system evolves.
Proof of the existence of a black hole in Arp 220 (or otherwise) is
then important in our understanding of the possible relationship
between quasars and ULIRGs.

\subsection{The supernova rate in Arp 220}

18 cm VLBI (3 $\times$ 8 milliarcsec resolution) continuum imaging
of Arp 220 has revealed the existence of more than a dozen sources
with 0.2$-$1.2 mJy fluxes (Smith et al. 1998), mostly in the western
nucleus. These compact radio sources were interpreted as supernova
remnants. This interpretation is consistent with a simple starburst
model for the IR luminosity of Arp 220 (Smith et al. 1998b), having
a constant SFR in the range 50$-$100 M$_\odot$ yr$^{-1}$, and a
supernova explosion rate in the range ${\cal R} \sim 1.75-3.5$
yr$^{-1}$.\footnote{The webpages of the Arecibo observatory further
report that in November 2002, a new VLBI experiment was conducted by
Lonsdale et al. and a preliminary continuum image has resulted in
the detection of roughly 30 supernova remnants candidates in Arp
220, about 10 of which lie in the eastern nucleus. This would be
direct evidence that intense star formation is occurring in both
nuclei, and not just the western one. } Smith et al. (1998) suggest
the adoption of a supernova explosion rate of 2 yr$^{-1}$, with an
uncertainty that could make it be twice this value. A radio
supernova would thus appear in Arp 220 at least once every six
months, and several individual SNRs would be visible at any given
moment.\footnote{A 2001 conference report by Lonsdale et al., while
confirming that the previously referred radio sources are indeed
supernovae, suggest that the explosion rate could be smaller than
the previous estimate. Apparently, there is yet no published report
after the 2002 observations.}

A model of the hidden nucleus was constructed using {\sc
starburst99} (Shioya, Trentham \& Tanigushi 2001) for which the star
formation rate derived was 267 M$_\odot$ yr$^{-1}$;  160 M$_\odot$
yr$^{-1}$ [107 M$_\odot$ yr$^{-1}$] of which correspond only to the
western [eastern] extreme starburst. For equal assumptions on the
IMF slope, the lower and upper limits on star masses, and the mass
needed for a star to evolve to a supernova, as compared with Smith
et al.'s (1998) work, a supernova rate of $\sim 4$ yr$^{-1}$ is
derived using this model, which is consistent with, but at the upper
end of, previous estimates.

Van Buren and Greenhouse (1994) developed, starting from Chevalier's
(1982) model for radio emission from supenova blast waves expanding
into the ejecta of their precursor stars, a direct relationship
between the FIR luminosity and the rate of supernova explosions. The
result is ${\cal R} = 2.3 \times 10^{-12} L_{\rm FIR}/L_\odot$
yr$^{-1}$.  They proved that the supernova rate resulting from this
relation was consistent with that derived from the star formation
rates in M82, NGC 253, and other galaxies. In the case of  ULIRGs,
Manucci et al. (2003) derived a similar expression. The latter
authors found, by studying a sample of 46 LIRGs and detecting 4
supernovae, that the supernova rate can be approximately given by
${\cal R} = (2.4\pm 0.1) \times 10^{-12} L_{\rm FIR}/L_\odot$
yr$^{-1}$, in nice agreement with Van Buren and Grennhouse's
results. Mattila and Meikle (2001) have also obtained a similar
value for the proportionality factor.

In the case of Arp 220, the total so-computed supernova rate is
${\cal R}= 2.8 \pm 0.1$ yr$^{-1}$, which is compatible with previous
results. The mentioned relationship  between $L_{\rm FIR}$ and
${\cal R}$ gives then the possibility of distributing the Arp 220
total supernova rate into the different components (i.e., disk,
western and eastern nuclei) according to their weight in the FIR
emission, and this is the approach followed here. As shown below,
this rate, together with the measured geometry of the system, fixes
the primary injection proton distribution. As compared with Local
Group Galaxies, the supernova rate in Arp 220 is $\sim 300$ times
larger (e.g., see the compilation produced by Pavlidou and Fields
2001, where the maximum rate occurs for M31, and it is 0.9
explosions per century).

\subsection{Dust emission}

The continuum emission from Arp 220, at wavelengths between $\sim 1$
cm and $\sim 10$ microns, was measured by Woody et al. (1989), Eales
et al. (1989), Scoville et al. (1991), Carico et al. (1992), and
Rigopoulou (1996), among others. These observations did not
distinguish, due to angular resolution, the different geometrical
components described in Figure 1, and were fitted with different
models for dust emission. In particular, Scoville et al. (1991)
already found that the continuum emission was mainly produced
thermally, by dust, and thus that it could be modelled with a
spectrum having an emissivity law $\nu^{\sigma}B(\epsilon, T)$.
Later, already with arcsec imaging, Scoville et al. (1997), Downes
and Solomon (1998), and Soifer et al. (1999) distinguished the
contribution of the two extreme starburst regions, and obtained
results compatible with previous measurements. However, the dust
emission modelling is strongly dependent on sizes, temperatures, and
emissivity indices of each of the emission regions, so that for a
small variation in any of these parameters, large changes in the
predicted fluxes of the components may result. This produces a
modelling degeneracy, acknowledged already by Soifer et al. (1999).
They provide a multicomponent fit for the dust emission of Arp 220,
and several possible scenarios, all compatible with observations,
were presented. These scenarios were recently re-analyzed by
Gonzalez-Alfonso et al. (2004), on the light of ISO-LWS
observations.

\begin{figure*}[t]
\centering
\includegraphics[width=.4\textwidth]{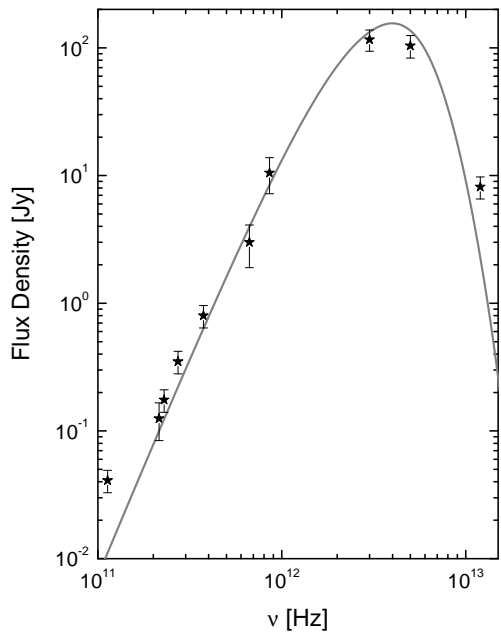}\hspace{0.5cm}
\caption{Data points and dust emission model assumed in this paper
for the IR-FIR radiation from Arp 220. Data points come from
literature quoted in the text, with typical errors of $\sim$20\%.
The theoretical curve is based on the assumption that the whole
IR-FIR luminosity is produced by dust located at each of the
components, emitting with a single temperature (42.2 K) and
emissivity index (1.5).} \label{dust}
\end{figure*}

Entering into too many details to represent the dust emission would
increase the number of parameters without a way of distinguishing
between different models with data now at hand. In addition, since
forthcoming $\gamma$-ray missions telescopes will not resolve the
different components, it is not really possible to relate subarcsec
FIR modelling with arcmin $\gamma$-ray observations. In the spirit
of Scoville et al. (1991), the simplest possible scenario is herein
adopted; i.e., the FIR emission is produced by dust in each of the
components, and that it is radiated with a single temperature and
emissivity law. The model (sum of the three contributions) derived
to fit the data ($\sigma=1.5$, $T=42.2$ K, see Appendix for details)
provides an excellent description of the observations, as can be
seen in Figure \ref{dust}. Note that, if anything, this model may
underestimates slightly what would be the real photon density,
particularly in the molecular disk, what implies that this
computation will not overestimate the inverse Compton contribution.
In any case, at high energies, in the dense environment of Arp 220,
inverse Compton emission is sub-dominant as compared with pion decay
$\gamma$-rays (see below).



Note that there are two points that deviates from the theoretical
curve in Figure \ref{dust}. The first is at the lowest frequencies,
where the dust emission model predicts less emission than observed.
Indeed, this behavior is correct, since at that frequency there is a
non-negligible non-thermal contribution coming from synchrotron
radiation as well as a thermal contribution coming from thermal
bremsstrahlung, computed below. This makes for this difference in
the fit. At the highest frequencies, the dust emission predicts less
emission than observed too, which is also correct, since at high
frequencies the source is optically thinner and better described by
a blackbody.

\section{Diffusion-loss equation}

The general diffusion-loss equation is given by (see, e.g., Longair
1994, p. 279; Ginzburg \& Syrovatskii 1964, p. 296) \be - D
\bigtriangledown ^2 N(E)+\frac{N(E)}{\tau(E)} - \frac{d}{dE} \left[
b(E) N(E) \right] - Q(E) = - \frac{\partial N(E)}{\partial t}.
\label{DL} \ee In this equation, $D$ is the scalar diffusion
coefficient, $Q(E)$ represents the source term appropriate to the
production of particles with energy $E$, $\tau(E)$ stands for the
confinement   timescale, $N(E)$ is the distribution of particles
with energies in the range $E$ and $E+dE$ per unit volume (see Table
\ref{units} for units), and $b(E)=-\left( {dE}/{dt} \right)$ is the
rate of loss of energy. The functions $b(E)$, $\tau(E)$, and $Q(E)$
will then be different depending on the nature of the particles
(i.e., electrons -- positrons, and protons, are subject to different
kind of losses and are also produced differently), but the form of
the equation will be the same for both. Here, two terms are to be
neglected: in the steady state,  $ {\partial N(E)}/{\partial t} =0,$
and the spatial dependence is considered to be irrelevant, so that $
D \bigtriangledown ^2 N(E) =0.$ This is reasonably under the
assumption of a homogeneous distribution of sources.

Eq. (\ref{DL}) can be --formally-- solved, as  can be proven by
direct differentiation,  by using the Green function  \be
G(E,{E^\prime})= \frac {1}{b(E)} \exp \left( -\int_E^{E^\prime} dy
\frac{1}{\tau(y) b(y)} \right), \ee such that for any given source
function, or emissivity, $Q(E)$, the solution is \be N(E) =
\int_E^{E_{\rm max}} d{E^\prime} Q({E^\prime}) G(E,{E^\prime}).
\label{SOL-DL}\ee Note that the integral in $E^\prime$ is made on
the primary energies which, after losses, produce secondaries with
energy $E$. In general, however, $G(E,{E^\prime})$ has not a close
analytical expression, and neither does $N(E)$. Numerical
integration techniques are then needed to compute Eq.
(\ref{SOL-DL}).

Instead of directly assuming a {\em steady state particle
distribution}, it is considered that the latter is the result of an
{\em injection distribution} being subject to losses and,
eventually, to secondary production, in the ISM. In general, the
injection distribution may be defined to a lesser degree of
uncertainty when compared with the steady state one, since the
former can be directly linked to observations, e.g., to the
supernova explosion rate. Such evolution of the injection spectrum
will be given as a solution of Eq. (\ref{DL}), with appropriate
$b(E)$ and $\tau(E)$ functions.

\begin{figure*}[t]
\centering
\includegraphics[width=.4\textwidth]{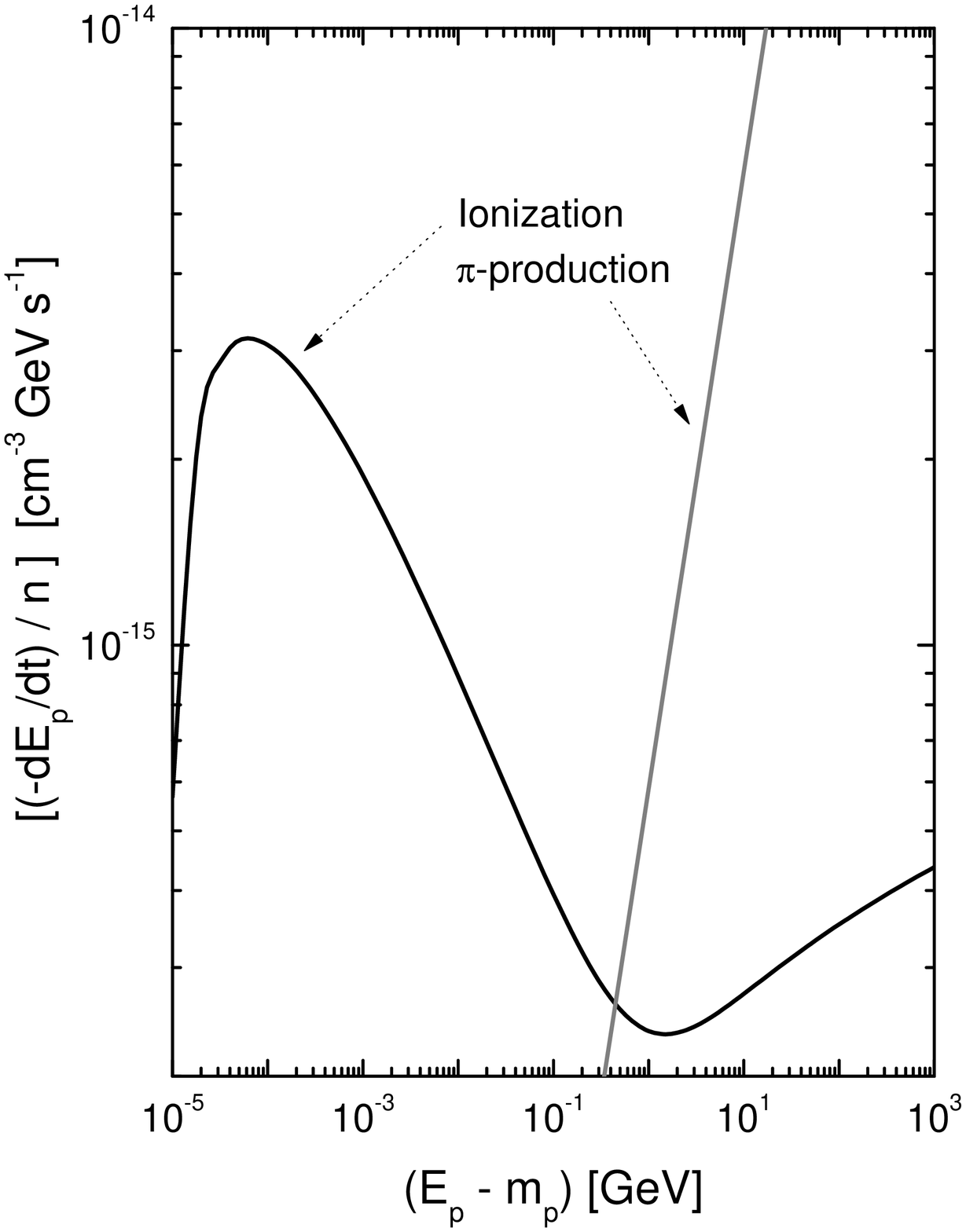}\hspace{0.5cm}
\includegraphics[width=.4\textwidth]{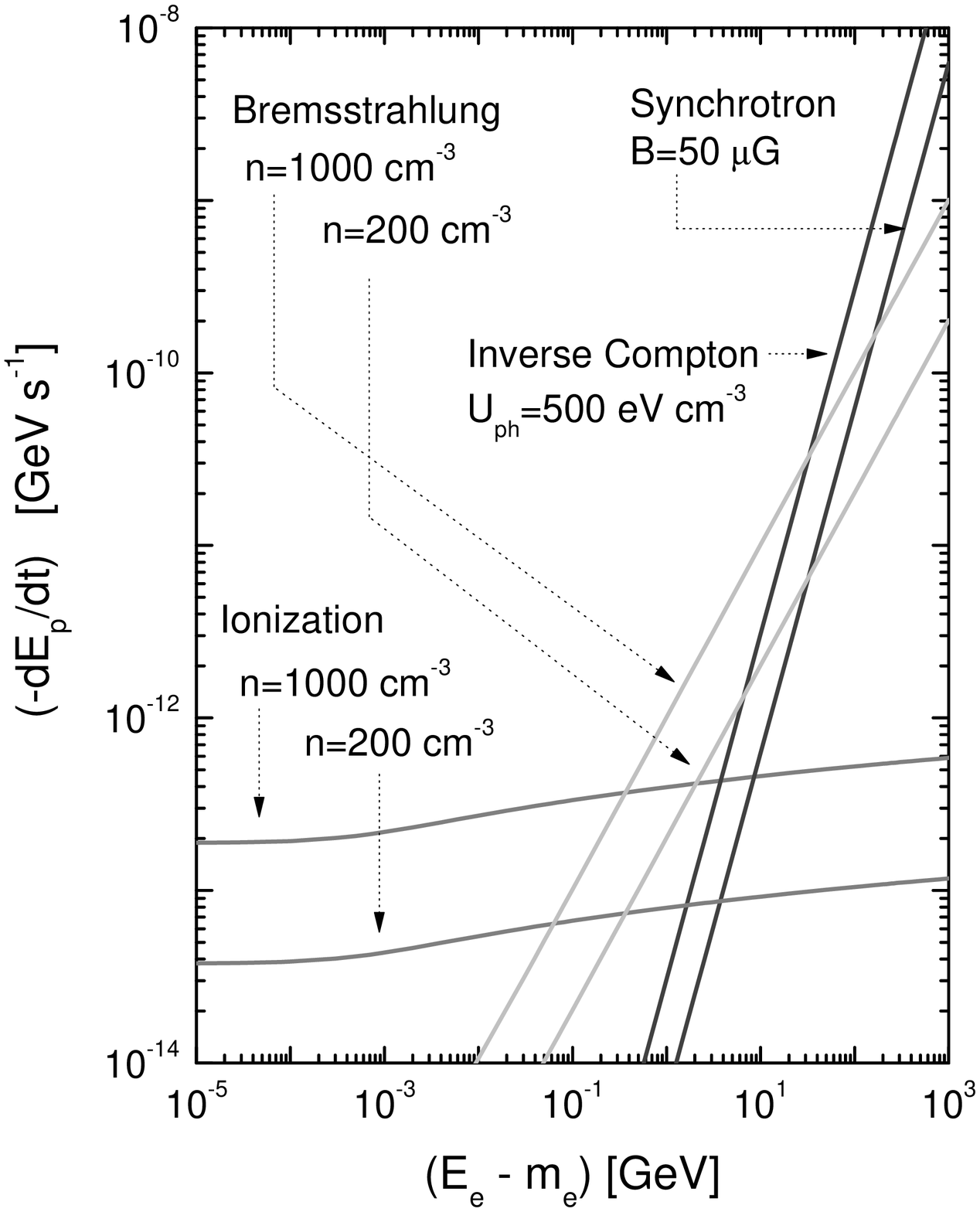}
\caption{Example of the rate of energy loss for protons (left panel)
and electrons (right panel) considered in this work. Protons losses
are mainly produced by ionization and pion production. Both are
proportional to the medium density,  and this is factored out (in
units of cm$^{-3}$). Electrons losses correspond to synchrotron and
bremsstrahlung radiation, inverse Compton scattering, and
ionization. A set of random parameters is assumed for this example
--shown in the figure--, additionally to the assumption that the
average density of the photon target is $\bar \epsilon = 1$ eV.  }
\label{rate}
\end{figure*}

The total rate of energy loss herein considered for protons is given
by the sum of that involving ionization and pion decay, as it is
discussed in the Appendix.\footnote{Pion decay losses are actually
catastrophic, since the inelasticity of the collision is about 50\%,
i.e., the beam particle looses an average 1/2 of its energy in every
interaction. Then, differently to ionization losses, pion production
could effectively remove particles from phase space. This effect,
however, is important when the proton population is described with
an steep spectrum, which is not the case of this work. One way to
treat catastrophic losses is to incorporate their
time-loss scale into the diffusion equation, as a term of the form
$N(E)/t_c$, where $t_c^{-1}=(dE/dt)/E$ (see, e.g., Mannheim \&
Schlickeiser 1994). By doing this, computing the steady spectrum
of protons, and comparing it with the result of using $(dE/dt)_{\rm
pion}$ as part of the `continuous losses' term, it is verified that
the difference is negligible and can be taken care of, if needed, by
a slight change in other parameter of the model. We then consider
for simplicity that pion losses are part of $b(E)$.}
An example of these rates of energy loss is shown in Figure
\ref{rate} (left panel). For electrons, the total rate of energy
loss considered is given by the sum of those involving ionization,
inverse Compton scattering, bremsstrahlung, and synchrotron
radiation, as it is also discussed in the Appendix. These rates of
energy loss are shown in the right panel of Figure \ref{rate} for a
particular choice of system parameters. In that figure, the inverse
Compton losses are computed in the Thomson approximation. The full
Klein-Nishina cross section is used while computing photon emission,
and either Thomson or extreme Klein-Nishina approximations, as
needed, are used while computing losses. This approach proves to be
accurate, while significantly reduces the computational time.

%
%

The confinement timescale will be given by the characteristic escape
time in the homogeneous diffusion model (Berezinskii et al. 1992, p.
50-52 and 78) $ \tau_D= {R^2}/( {2D(E)}) ={\tau_0}/( \beta (E/{\rm
GeV})^{\mu} ), \label{T-P0} $ where $\beta$ is the velocity of the
particle in units of $c$, $R$ is the spatial extent of the region
from where particles diffuse away, and $D(E)$ is the
energy-dependent diffusion coefficient, whose dependence is assumed
$\propto E^{\mu}$, with $\mu \sim 0.5$. $\tau_0$ is the
characteristic escape time at $\sim$ 1 GeV. Note that, whereas the
form of $\tau_D$ is assumed the same for both protons and electrons,
its value at a fixed energy is only the same for particles with
equal Lorentz factors (and thus equal $\beta$).
The total escape timescale will also take into account that
particles can be carried away by the collective effect of stellar
winds and supernovae. In general, it is reasonable to suppose that
this timescale ($\tau_c$) is within one order of magnitude of
$\tau_0$. $\tau_c$ is indeed $\sim R / V$, where $V$ is the
collective wind velocity. Thus, in general, $ \tau^{-1}(E_p)=
\tau_o^{-1} \beta \left({E}/{{\rm GeV}}\right)^{\mu} +
{\tau_c}^{-1}. \label{T-P} $
%

Note that if $Q(E)$ is a power
law, $N(E)$ scales linearly with its normalization. However, there
is no immediate scaling property with the density of the ISM,
which enters differently into the several expressions of losses
that conform $b(E)$.


\section{Computation of secondaries  \label{SEC-SEC}}

For the production of secondary electrons, only knock-on and pion
processes are taken into account. These processes dominate by more
than an order of magnitude the production of electrons at low and
high energies, respectively, when compared with neutron beta decay
(see, e.g., Marscher \& Brown 1978 and Morfill 1982 for discussions
on this issue).

\subsection{Electrons from knock-on (or Coulomb) interactions}

Knock-on (or Coulomb) collisions are interactions in which the
proton CR transfers an energy far in excess of the typical binding
energy of atomic electrons, so producing low-energy relativistic
electrons. The cross section for knock-on production was calculated
by Bhaba (1938) and subsequently analyzed by Brunstein (1965) and
Abraham et al. (1966), among others. The differential probability
for the production of an electron of energy $E_e$ and corresponding
Lorentz factor $\Gamma_e=E_e/m_e c^2$, within an interval $(\Gamma_e
- d\Gamma_e, \Gamma_e + d\Gamma_e)$, produced by the collision
between a CR of particle species $j$, and energy factor $\Gamma_j$
and a target of charge $Z_i$ and atomic number $A_i$ is, in units of
grammage, \be \Phi (\Gamma_e, \Gamma_j) d\Gamma_e = \left[ \frac{2
\pi N_0 Z_i r_e^2 Z_j^2} {A_i (1-\Gamma_j^{-2}) } \left( \frac{1}
{(\Gamma_e - 1)^2} - \frac{s \left( \Gamma_j +
\frac{s^2+1}{2s}\right)} {(\Gamma_e - 1) \Gamma_j^2} + \frac{s^2}{2
\Gamma_j^2} \right) \right]\, d\Gamma_e \; {\rm cm}^2 \; {\rm
g}^{-1}. \ee Here $N_0$ is the Avogadro's number,
$r_e=e^2/mc^2=2.82\times 10^{-13}$ cm is the classical radius of the
electron, and $s=m_e / (A_i m_p)\sim 1/1836$ (see below). Note that
the probability for interaction is proportional to $Z_i / A_i$.
Then, it will suffice to assume that the interstellar medium is 90\%
hydrogen and 10\% Helium and neglect the contribution of higher
atomic numbers. This approximation introduces negligible error.
Contributions by various nuclei in the colliding CR population are
more important, since the probability for interaction is
proportional to $Z_j^2$. If the total contribution of all primaries
with charge $Z \geq 2$ relative to that of protons is $\sim 0.75$,
then $
 \sum_i  \sum_j  \Phi( \Gamma_e, \Gamma_j)
\sim 1.75\, \Phi ( \Gamma_e, \Gamma_p). $

The maximum transferable energy in this kind of collisions is (e.g.,
Abraham et al. 1966) $ \Gamma_{\rm max} = 1 + \left( \Gamma_p^2 -1
\right) / \{ s \left( \Gamma_p + [{s^2+1}/{2s}] \right) \}. $ Thus,
the maximum possible energy is limited only by the maximum value of
$\Gamma_p$, while the minimum proton Lorentz factor that is needed
to generate an electron of energy $E_e$ is fixed by solving the
inequality $\Gamma_e \leq \Gamma_{\rm max}$. The result is that
$\Gamma_p \geq \Gamma_1$, with $ \Gamma_1 =  [1/2] s (\Gamma_e -1 )
+ \sqrt{ 1+ \frac 12 (1+s^2)(\Gamma_e-1)+ \frac 14 s^2 (\Gamma_e
-1)^2  }. $ With this in mind, the source function for knock-on
electrons to be considered in the diffusion-loss equation is then
given by \be Q_{\rm knock}(E_e) \sim 1.75 \; m_p \; n \;4\pi \;
\int_{E_{1,p}} \Phi(E_e,E_p)\, J_p(E_p) \, dE_p \; ,
\label{knocsource}\ee where $E_{1,p}=\Gamma_1 \, m_p$,
$\Phi(E_e,E_p)=\Phi(\Gamma_e,\Gamma_p)/m_e$, i.e. energies, instead
of Lorentz factors, are used to write the final integral, and $J_p$
is the CR proton intensity
($J_p(E)=(c\beta/4\pi)N(E)$).\footnote{Note that for the computation
of secondary electrons, sometimes it is more convenient to use
$Q(\Gamma)$, the emissivity as a function of the electron Lorentz
factor, instead of $Q(E)$. They are related by $Q(\Gamma)
d\Gamma=Q(E) dE$, then $Q(\Gamma)$-units are cm$^{-2}$ s$^{-1}$
sr$^{-1}$ unit-$\Gamma^{-1}$. In order to convert electron and
positron emissivities expressed as a function of $\Gamma$ to those
expressed as a function of energy, which are those entering into the
expression of the diffusion-loss equation adopted, one has then to
divide by the electron mass. Note also that the equality
$J_p(\Gamma) d\Gamma= J_p(E) dE$ holds. Similarly, the relationship
between $\Phi(E_e,E_p)$ and $\Phi(\Gamma_e,\Gamma_p)$ can be
obtained. }

\begin{figure*}[t]
\centering
\includegraphics[width=.4\textwidth]{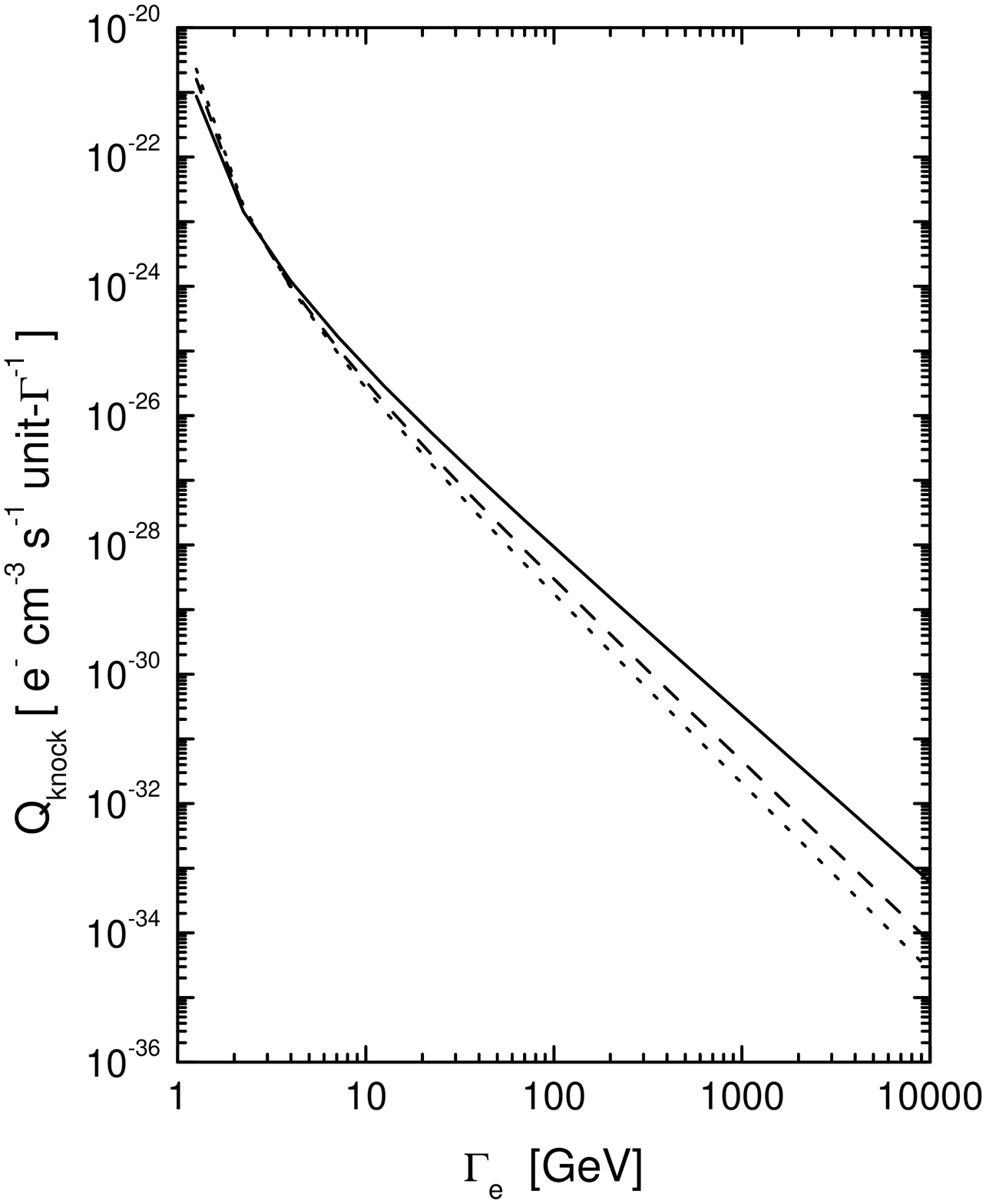}
\hspace{1cm}
\includegraphics[width=.4\textwidth]{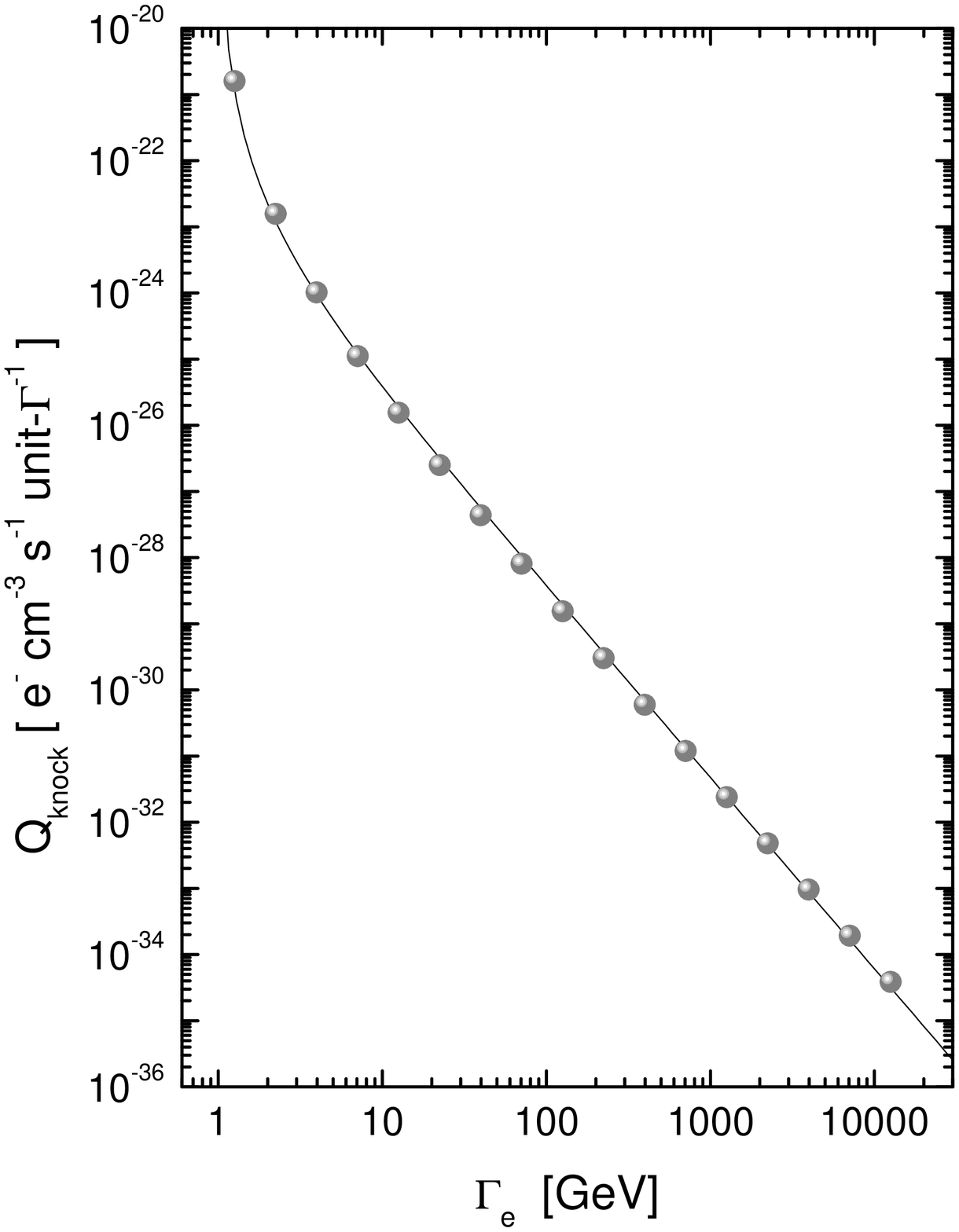}
\caption{Left: Knock-on source function for different CR intensity
$J_p(E_p)=A (E_{\rm kin}/{\rm GeV})^{\alpha}$ protons cm$^{-2}$
s$^{-1}$ sr$^{-1}$ GeV$^{-1}$. We have normalized the source
function by taking an ISM density ($n=1$ cm$^{-3}$) and unit
normalization of the incident proton spectrum, $A$=1. Curves shown
are, from top to bottom, the corresponding to $\alpha=-2.1, -2.5$,
and $-2.7$. Right: Simple power law fit of the knock-on source
function for $\alpha=-2.5$. Similar fits can be plotted for all
values of $\alpha$, see text for details. } \label{knock-fig}
\end{figure*}

If the CR intensity is described by a power law whose exponent is
exactly an integer or half of an integer, i.e., $-2, -2.5, -3,$
etc., lengthy analytical expressions for the knock-on source
function can be obtained. This is no longer true for generic power
laws. Examples of the results for the computation of the knock-on
source function are given in Figure \ref{knock-fig}.
As it is shown there, and was first proposed by Abraham et al.
(1966), the behavior of the knock-on source function can be well
represented by a power law of the form $ Q_{\rm knock}(E_e) \sim
constant \times (\Gamma_e - 1)^{-b} {\rm electrons} \; {\rm cm}^{-3}
\; {\rm s}^{-1}\; {\rm GeV}^{-1}. $ An example of such a description
can be found in the right panel of Figure \ref{knock-fig}, where the
spectrum obtained using Eq. (\ref{knocsource}) is superposed to the
fit.

\subsection{$\gamma$-rays from neutral pion decays}

The $\pi^0$ emissivity resulting from an isotropic intensity of
protons, $J_p(E_p)$, interacting with --fixed target-- nuclei with
number density $n$, through the reaction $p+p \rightarrow p+\pi^0
\rightarrow p+2\gamma $, is given by (e.g., Stecker 1971) \be
Q_{\pi^0}(E_{\pi^0}) = 4 \pi n \int_{E_{th}(E_{\pi^0})} dE_p\,
J_p(E_p) {d\sigma(E_{\pi^0}, E_p)\over dE_{\pi^0}}\;, \ee where
$E_{p}(E_{\pi^0})$ is the minimum proton energy required to
produce a pion with total energy $E_{\pi^0}$, and is determined
through kinematical considerations.
%
%
${d\sigma(E_{\pi^0}, E_p) / dE_{\pi^0}}$ is the differential cross
section for the production of a pion with energy $E_{\pi^0}$, in the
lab frame, due to a collision of a proton of energy $E_p$ with a
hydrogen atom at rest.
The $\gamma$-ray emissivity is obtained from the neutral pion
emissivity $Q_{\pi^0}$ as \be \label{pion-prog}
{Q_\gamma(E_\gamma)}_{\pi} = 2 \int_{E_{\pi^0}^{min} (E_\gamma)}
dE_{\pi^0} {Q_{\pi^0}(E_{\pi^0}) \over (E_{\pi^0}^2 - m_{\pi^0}^2
c^4)^{1/2}} \ee
where $E_{\pi^0}^{min} (E_\gamma) = E_\gamma +
m_{\pi^0}^2 c^4 / (4E_\gamma)$ is the minimum pion energy required
to produce a photon of energy $E_\gamma$ (e.g., Stecker 1971).

Recently, Blattnig et al. (2000) developed
parameterizations of the differential cross sections regulating the
production of neutral and charged pions.  On one hand, Blattnig et
al. have presented a parameterization of the Stephens and Badhwar's
(1981) model by numerically integrating the Lorentz-invariant
differential cross section (LIDCS).\footnote{The invariant
single-particle distribution is defined by $ f(AB \rightarrow
CX)\equiv E_c\frac{d^{3}\sigma}{d^{3}{ p}_{c}}\equiv
E\frac{d^3\sigma}{d^3p} =\frac{E}{p^2}\frac{d^{3}\sigma}{dp
d\Omega}$ where ${d^{3}\sigma}/{d^{3}{p}_{c}}$ is the differential
cross-section (i.e. the probability per unit incident flux) for
detecting a particle $C$ within the phase-space volume element
$d^{3}{p}_{c}$. $A$ and $B$ are the initial colliding particles, $C$
is the produced particle of interest, and $X$ represents all other
particles produced in the collision. $E$ is the total energy of the
produced particle $C$, and $\Omega$ is the solid angle.
 This quantity is invariant under Lorentz
transformations and is called LIDCS. LIDCSs for inclusive pion
production in proton-proton collisions contain dependences on the
energy of the colliding protons (through the energy of the center of
mass in the collision $\sqrt{s}$), on the energy of the produced
pion (whose kinetic energy is $T_{\pi}$), and on the scattering
angle of the pion ($\theta$).  Total cross sections, $\sigma$, which
depend only on $\sqrt{s}$, and spectral (or differential) cross
sections, ${d\sigma}/{dE}$, which depend on $\sqrt{s}$ and
$T_{\pi}$, can be extracted from the LIDCS by integration.  If
azimuthal symmetry is assumed, these cross sections are $
\frac{d\sigma}{dE}  =  2\pi p \int_0^{\theta_{max}} d\theta
E\frac{d^{3}\sigma}{d^{3}{p}} \sin\theta, $ and $ \sigma = 2\pi
\int_0^{\theta_{max}} d\theta \int_{p_{min}}^{p_{max}} dp
E\frac{d^{3}\sigma}{d^{3}{p}} \frac{p^{2}\sin\theta}{\sqrt{p^{2} +
m_{\pi}^2}}, $ where $\theta_{max}$, $p_{max}$, and $p_{min}$ are
the extrema of the scattering angle and momentum of the pion
respectively, and $m_{\pi}$ is the rest mass of the pion. These
extrema are determined by kinematic considerations (see Blattnig et
al. 2000 for details). Then, starting from different LIDCS
parameterizations it is possible to integrate these over the
kinematics to obtain the corresponding parameterizations for the
total and differential cross sections. The accuracy of the latter
forms will solely depend on the accuracy of the parameterizations of
the LIDCS. } The expression of such parameterization is divided into
two regions, depending on the (laboratory frame) proton energy
(Blattnig et al. 2000).
\begin{figure*}[t]
\centering
\includegraphics[width=.4\textwidth]{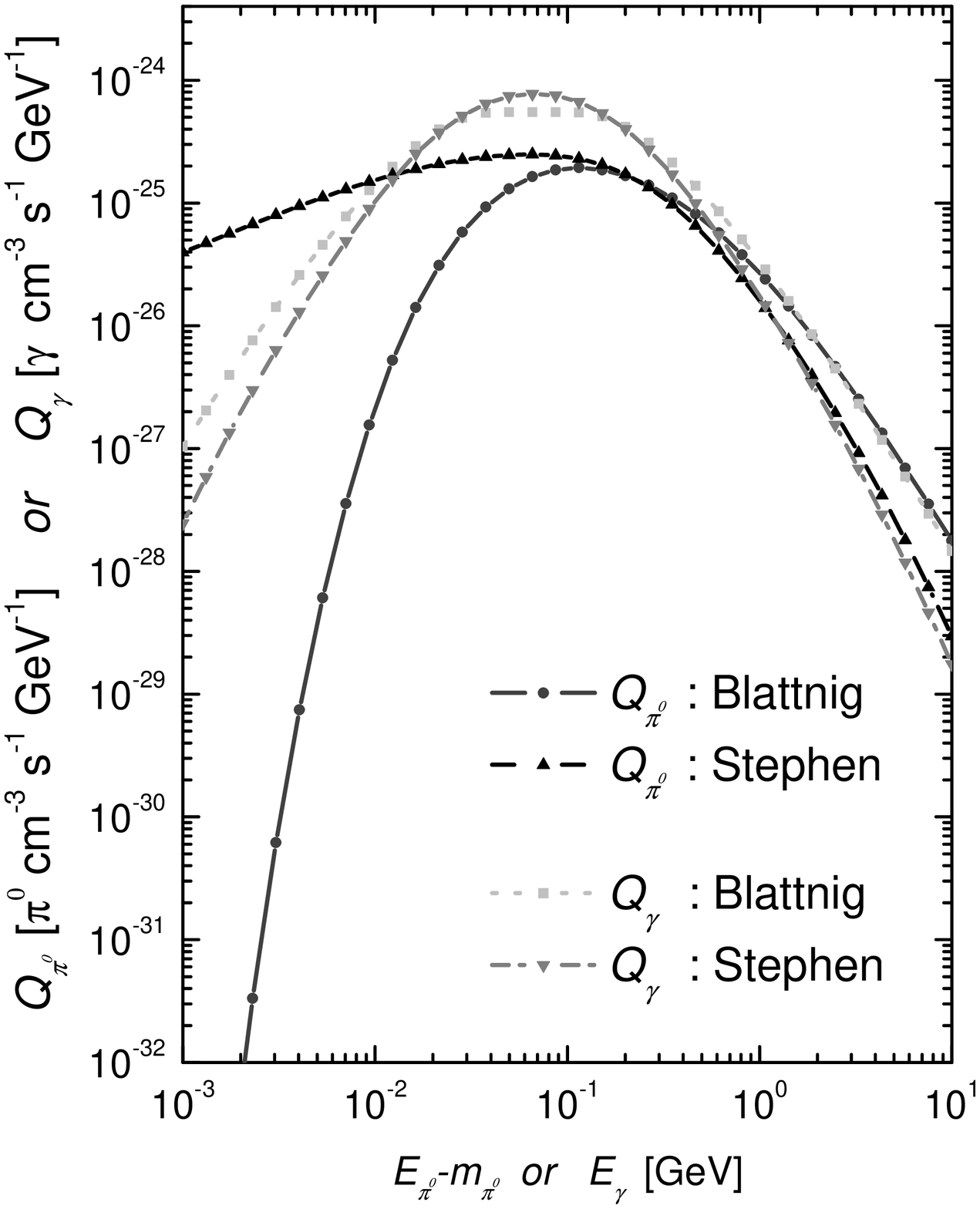}
\hspace{1cm}
\includegraphics[width=.4\textwidth]{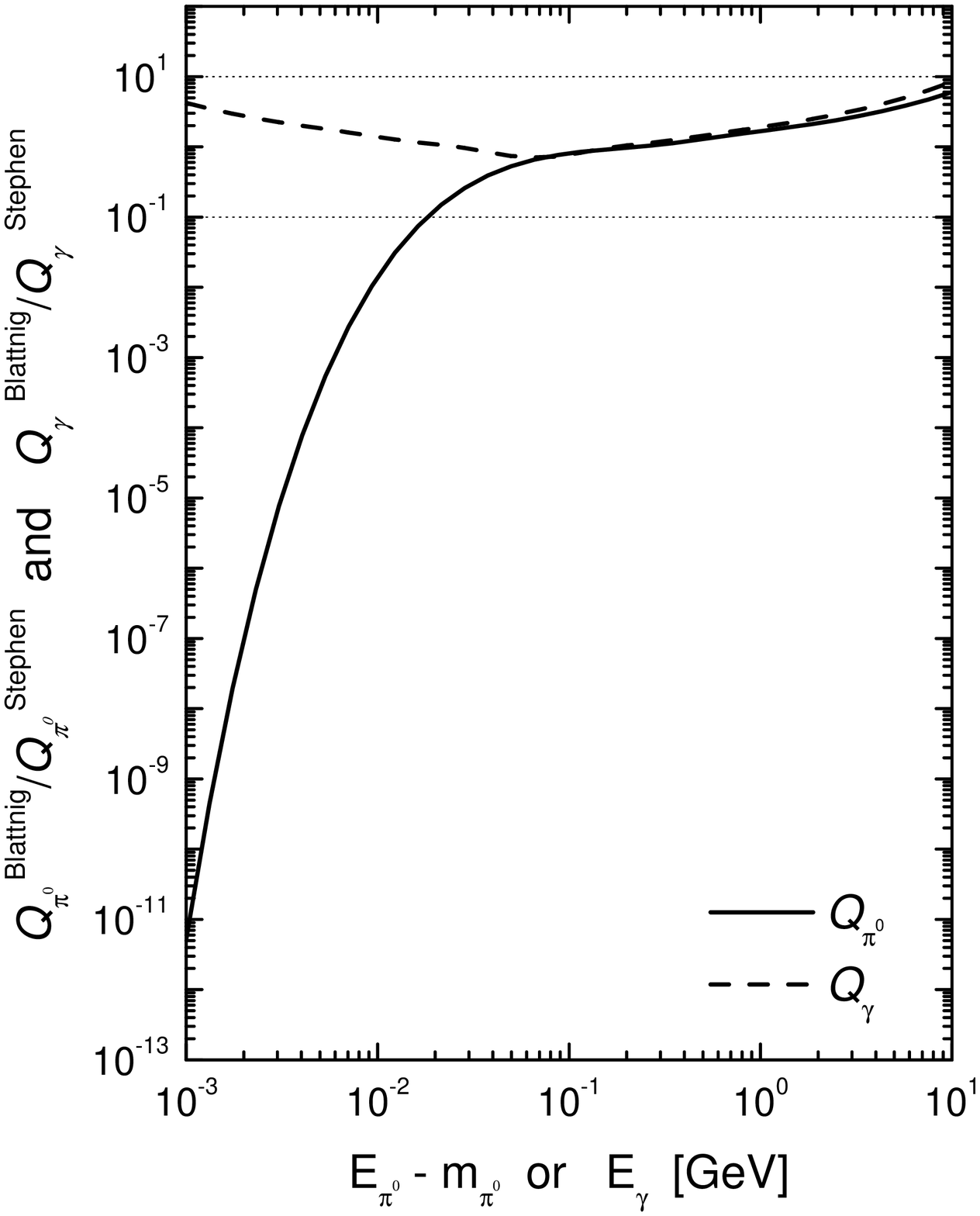}
\caption{Left: $\pi ^0$ and $\gamma$-emissivities computed using
Blattnig et al.'s (2000) and Stephen and Badhwar's (1981)
parameterizations. Right: Discrepancies between cross section models
are shown as the ratio of the emissivities of secondary neutral
particles. In both panels,  $n=1$ cm$^{-3}$, and an Earth-like
proton spectrum ($\propto E^{-2.75}$) are assumed.}
\label{pionfigure1}
\end{figure*}
On the other hand, Blattnig et al.'s new parameterization has,
particularly in the case of neutral pions, a much simpler analytical
form. It is given by \be
    \frac{d\sigma(E_{\pi^0}, E_p)}{dE_{\rm \pi^0}}
    =10^{-27} e^{\left(
    -5.8-{1.82}/{(E_p-m_p)^{0.4}}
     +{13.5}/{(E_{\rm \pi^0}-m_{\rm \pi^0})^{0.2}}
     -{4.5}/{(E_{\rm \pi^0}-m_{\rm \pi^0})^{0.4}} \right)}
     {\rm cm}^{2}\;      {\rm GeV}^{-1}
     \label{param-neutral}
     \ee
which ease the computation of the pion spectrum as compared to the
isobaric (Stecker 1971) or scaling models (Stephens \& Badwhars
1981), see, e.g. Dermer (1986), although yet requiring numerical
integration subroutines. [Recall that rest masses and energies must
be given, in the last equation, in units of GeV.]
Blattnig et al.'s parameterization were not yet applied to compute
$\gamma$-ray emission. Then, a brief analysis can prove useful.
Specifically, the computed pion decay emissivity using the new
Blattnig et al.'s (2000) parameterization (Eq. \ref{param-neutral})
is herein compared with that corresponding to the Stephen and
Badhwar's (1981) one,  assuming the same proton injection and
density as in Dermer (1986).\footnote{The proton spectrum is the
Earth-like one, $J_p(E_p)=2.2\,E_p^{-2.75}$ protons cm$^{-2}$
s$^{-1}$ sr$^{-1}$ GeV$^{-1}$ and $n=1$ cm$^{-3}$. The resulting
$\gamma$-ray emissivity is multiplied by 1.45 to give account of the
contribution to the pion spectrum produced in interactions with
heavier nuclei (Dermer 1986). The maximum proton energy is assumed
as 10 TeV.}

Using Eq. (\ref{pion-prog}), it is possible to see that under the
Blattnig et al. new parameterization, the number of pions produced
at low ($E_{\pi^0}-m_{\rm \pi^0}<10^{-2}$ GeV) energies is
significantly less than that produced using the alternative model.
Fig. 6 of Blattnig et al. (2000) shows that their new differential
cross section parameterization decreases rapidly at low energies and
goes to approximately zero at 10 MeV. Fig. 5 of the same paper shows
that Stephen and Badhwar's cross section, instead, is much larger at
very low pion energies (see Blattnig et al. 2000b for further
details). Noteworthy, this fact, however, does not substantially
affect the $\gamma$-ray emission in the region of interest since to
produce a photon of energy $\sim$10$^{-2}$ GeV, pions of minimum
energy of $\sim 0.5$ GeV are required, and at these energies, the
pion spectrum using both approaches agrees reasonable well (i.e. the
$\gamma$-ray spectrum is within an order of magnitude at all
energies). This comparison is shown in detail in the two panels of
Figure~\ref{pionfigure1}.

Regrettably, it seems not possible to answer which parameterization
is the correct one at low energies with current experimental data
(see Blattnig et al. 2000 for a discussion). The problem being that
the shapes of the two spectral distributions, (${d\sigma(E_{\pi^0},
E_p)}/{dE_{\rm \pi^0}}$), look quite different even when both
original LIDCSs  have a similar fit to the data at low transverse
momentum of the produced pion, where the cross section is the
greatest, and that both integrate to the same total cross section.
Notwithstanding, at high transverse momentum, Stephen and Badwhars's
parameterization overpredicts the cross section for several orders
of magnitude, and Blattnig et al.'s form is preferred, (Eq.
\ref{param-neutral}). Then, for neutral pion decay computations, Eq.
\ref{param-neutral} is adopted in our computations. In the case of
charged pions, Badwhar's (1977) LIDCS is considered the most
reliable at all energies, and then their corresponding spectral
distributions are adopted, see below.


\subsection{Electrons and positrons from charged pion decay}

\begin{figure*}[t]
\centering
\includegraphics[width=.4\textwidth]{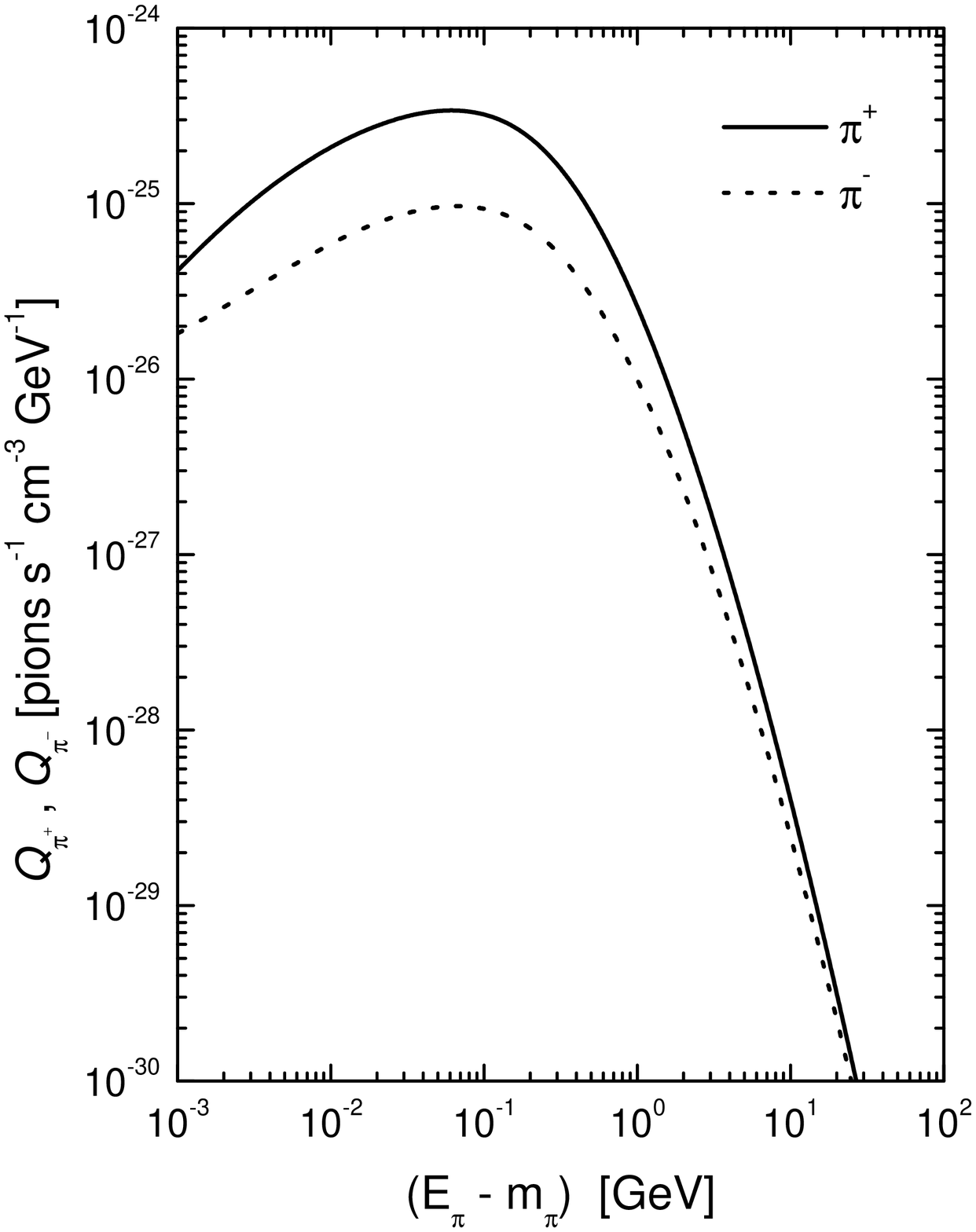}
\hspace{1cm}
\includegraphics[width=.4\textwidth]{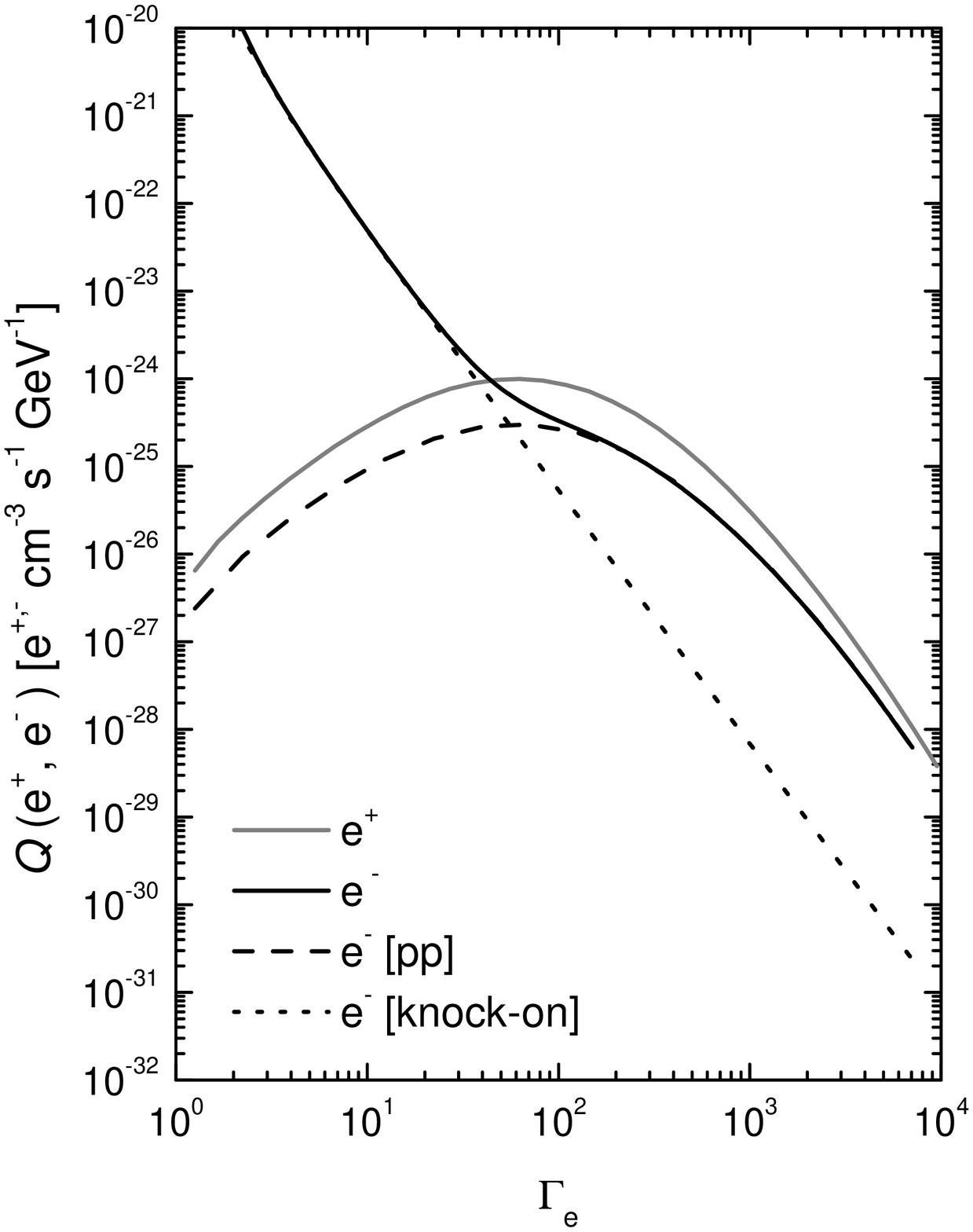}
\caption{Left: $\pi^\pm$-emissivities produced using Blattnig et
al.'s (2000) parameterizations of Bhadwar et al.'s (1977) spectral
distribution, ${d\sigma}/{dE}$. Right: $e^\pm$-emissivities.  In the
case of electrons, the total emissivity adds up that produced by
knock-on interactions, which dominates at low energies. In both
panels, $n=1$ cm$^{-3}$, and an Earth-like proton spectrum ($\propto
E^{-2.75}$) are assumed.} \label{pionfigure2}
\end{figure*}
%
%

Positron production occurs through muon decay in the reactions $ p+p
\rightarrow X+\pi^+$ with the pion then decaying as $
\pi^+\rightarrow \mu^+ + \nu_e + \bar \nu_\mu$. Electron production
occurs, similarly, through, $p+p \rightarrow X+\pi^-$ with the pion
then decaying as $ \pi^-\rightarrow \mu^- + \bar \nu_e + \nu_\mu$.
Considering first the latter decays in the frame at rest with the
pion, conservation of energy and momentum imply $ {p_\mu}^\prime =
\left( {{E_\mu}^2}^\prime - {m_\mu}^2 \right)^{1/2} = [{m_\pi}^2 -
{m_\mu}^2]/{2m_\pi},$ where $m_{\mu,\pi}$ are the masses of the muon
and pion, respectively, $E$ are total energies and the prime is used
to represent the pion rest frame. This implies that the energy of
the pion in such frame is $ {\Gamma_\mu}^\prime = {
{E_\mu}^\prime}/{m_\mu}  = [{m_\pi}^2 + {m_\mu}^2]/[2 m_\pi
m_\mu]\sim 1.04 .$ The value of $ {\Gamma_\mu}^\prime $ implies, as
long as the velocity of the pion in the laboratory frame is not
exceedingly small ($\Gamma_\pi
> 1.04$), that the muon is practically at rest in the rest frame
of the pion, and that as seen from the lab, $ \Gamma_\mu \sim
\Gamma_\pi$. Then, per unit Lorentz factor, the muon emissivity is
equal to that of the pion $ Q_{\pi^+}(\Gamma_{\pi^+})=
Q_{\mu^+}(\Gamma_{\mu^+})$, $Q_{\pi^-}(\Gamma_{\pi^-})=
Q_{\mu^-}(\Gamma_{\mu^-}). $ The charged pion emissivity resulting
from an isotropic distribution of protons $J_p(E_p)$ interacting
with --fixed target-- nuclei found with number density $n$ can be
computed as that of the neutral pions, by just changing the spectral
distribution \be Q_{\pi^{\pm}}(\Gamma_{\pi^{\pm}}) = 4 \pi n
\int_{\Gamma_{th}(\Gamma_{\pi^\pm})} d\Gamma_p\, J_p(\Gamma_p)
{d\sigma(\Gamma_{\pi^\pm}, \Gamma_p)\over d\Gamma_{\pi^\pm}}\;, \ee
where $\Gamma_{p}(\Gamma_{\pi^\pm})$ is the minimum proton Lorentz
factor required to produce a pion (either positively or negatively
charged) with Lorentz factor $\Gamma_{\pi^\pm}$. Thus, knowledge of
the spectral distribution ${d\sigma(\Gamma_{\pi^\pm}, \Gamma_p)/
d\Gamma_{\pi^\pm}}$ secures knowledge of the muon emissivity. Use of
the new parameterizations of the Bhadwar et al. (1977) LIDCS
(Blattnig et al. 2000). Figure \ref{pionfigure2}, left panel, shows
an example of the $\pi^+$-- and $\pi^-$--emissivity. The electron
and positron emissivities are computed as a three-body decay process
(see, e.g. Schlickeiser 2002, p. 115):\footnote{The three-body muon
decay is actually a simplification. In case of charged pion decays,
muons appear to be all polarized. This means that positrons are
mainly emitted forward while electrons are emitted backwards in the
CMS, and results in different distributions of these particles in
the laboratory system (e.g., see Moskalenko \& Strong 1998). }
\be Q_{e^{\pm}}(\Gamma_{e^{\pm}})= {\int_1}^{{\Gamma_e}^{\prime {\rm
max}}} d{\Gamma_e}^{\prime} \frac 12 \frac{P({\Gamma_e}^{\prime})}{
\sqrt{ {\Gamma_e}^{\prime 2} -1 } } \int_{\Gamma_{\mu
1}}^{\Gamma_{\mu 2}} d{\Gamma_\mu}
\frac{Q_{\mu^\pm}(\Gamma_{\mu^\pm})}{\sqrt{{\Gamma_\mu}^2-1}}. \ee
Here ${{\Gamma_e}^{\prime {\rm max}}}=104$, $ \Gamma_{\mu 1, \mu 2}
=  \Gamma_{e^{\pm}} {\Gamma_e}^{\prime} \mp \sqrt{{\Gamma_e}^{\prime
2}-1} \sqrt{{\Gamma_e}^{ 2}-1}, $ and the function $P$ is
$P({\Gamma_e}^{\prime})  =  2 {{\Gamma_e}^{\prime}}^2 \left[ 3 -
\frac{2{{\Gamma_e}^{\prime}}}{{{\Gamma_e}^{\prime {\rm max}}}}
\right] /{({{\Gamma_e}^{\prime {\rm max}}})^3}. $ Figure
\ref{pionfigure2},  right panel, shows an example of the $e^+$-- and
$e^-$--emissivity, as implemented in the code. These results are
compatible with previous computations.

\section{Steady distributions, emissivities, and magnetic fields
in Arp 220}

\subsection{Protons}

The injection proton emissivity is here, following Bell (1978),
assumed to be a power law in proton kinetic energies, with index $p$
(herein $p=2.2$), \be Q_{\rm inj}(E_{\rm p,\, kin}) = K
\left(\frac{E_{\rm p,\, kin}}{\rm
GeV}\right)^{-p} , 
\ee where $K$ is a normalization constant.\footnote{ This expression
is strictly valid for proton Lorentz factors much larger than 1.
However, it differs from the exact expression at very low energies,
Eq. (5) of Bell (1978), by less than a factor of 3, at most, what
produces an overall negligible difference. The spectrum of particles
accelerated by SNR is a power-law in rigidity (e.g.,
Ellison et al 2004).}
This normalization is to be obtained from the total power
transferred by supernovae into CRs kinetic energy within a given
volume \be \int_{E_{\rm p,\,kin,\, min}}^{E_{\rm p,\, kin,\, max}}
Q_{\rm inj}(E_{\rm p,\, kin}) E_{\rm p,\, kin} dE_{\rm p,\, kin} =
-K \frac{E_{\rm p,\,kin,\,min}^{-p+2}}{-p+2}
 \equiv \frac{ \sum_i \eta_i {\cal P} {\cal R}_i } {V}
\ee where it was assumed that $p\neq 2$, used the fact that $E_{\rm
p,\,kin\,min} \ll E_{\rm p,\,kin,max}$ in the second equality, and
defined ${\cal R}_i$ ($\sum_i {\cal R}_i={\cal R}$) as the rate of
supernova explosions in the star forming region being considered,
$V$ being its volume, that transfer a fraction $\eta_i$ of the
supernova explosion power (${\cal P} \sim 10^{51}$ erg) into CRs.
The summation over $i$ takes into account that not all supernovae
will transfer the same amount of power into CRs (alternatively, that
not all supernovae will release the same power). The rate of power
transfer is assumed to be in the range 0.05 $\lesssim \eta_i
\lesssim$ 0.25 (e.g., Torres et al. 2003 and references therein),
uniformly distributed. Then, taking a ten-piece histogram, $\sum_i
\eta_i {\cal R}_i=0.165 {\cal R}$.
%
%
%
Note that $E_{\rm p,\, kin, min}$ is also fixed by requiring that
the minimum kinetic proton energy with which a CR escapes from a
shock front be larger than $2m_p v_s^2$ (Bell 1978). For shock
velocities of the order of 10$^{3-4}$ km s$^{-1}$, this is in the
range of a few MeV. A value of 10 MeV is taken to fix numerical
constants, although its precise value is not a relevant parameter in
this problem.


These assumptions imply that the injection is fixed as \ba Q_{\rm
inj}(E_p) = \! \left[\frac{{\cal P} \times {\sum_i \eta_i {\cal
R}_i} \times V^{-1}}{\rm GeV \,{\rm s}^{-1}\, {\rm
cm}^{-3}}\right] \!\! [p-2] \!\! \left[ \frac{ E_{\rm p, \, kin,\,
min}}{\rm GeV}\right]^{p-2}\!\! \left[\frac{E_p-m_p}{\rm
GeV}\right]^{-p}\!\! {\rm GeV}^{-1}\, {\rm cm}^{-3} \, {\rm
s}^{-1} \label{INJ-P}.\ea
The numerical solution of the diffusion-loss equation for protons,
subject to the losses described in the Appendix, is shown in Figure
\ref{steadyarp}. It is assumed that the unknown diffusion timescale
is proportional to $\tau_o=1$ Myr for the extreme starburst regions,
and to $\tau_o=10$ Myr for the much larger volume occupied by the
disk. The chosen $\tau_o$ is a factor of 2--10 less than that
estimated for our Galaxy or M33 (e.g., Gaisser 1990, Duric et al.
1995), and parallels that obtained in the study of NGC 253 and M82,
which are a galaxies presenting a more similar environment to Arp
220 (Paglione et al. 1996, Blom et al. 1999). The shorter residence
timescale for the extreme starburst regions actually makes for a
conservative assumption: if erring, it would be (slightly)
underestimating the $\gamma$-ray flux. The previous values will
stand for the assumed model of Arp 220, but others are explored in
the middle panel of Figure \ref{steadyarp}. There, corresponding to
the western nucleus, the ratio between the proton distribution
obtained with $\tau_o=10, 0.5,$ and 0.1 Myr, and that obtained with
the adopted model in this paper ($\tau_o$=1 Myr), is shown. Unless
in the extreme case of very low $\tau_o$, the steady distribution is
not significantly sensitive to this parameter. Differences increase
with energy, and amount to less than 5\% at 1 TeV, what is
practically unobservable regarding the $\gamma$-ray output. The
latter can increase or decrease slightly, depending on the value of
$\tau_o$ adopted, and uncertainties in other parameters can wash out
this effect completely.
For the case with $\tau_o=0.1$ Myr, there would be a reduction of
the relativistic distribution of protons by a factor of 2 at the
highest energies. However, such a low value of $\tau_o$ is not
favored: it represents a residence time two orders of magnitude
shorter than that of our Galaxy, and it would largely dominate over
the radiative timescales, what is not the case in dense molecular
clouds (see, e.g., the appendix of the work by Marscher \& Brown
1978).

Ionization (pion) losses dominates at low (high) energy, and this
change in the dominant mechanism for the energy loss produces the
kink that appears in the curves of Figure \ref{steadyarp} around a
kinetic energy of 300 MeV. Note that the steady distribution in each
of the components is similar (and actually, slightly larger for the
extreme starburst regions) despite of their different sizes. This
implies that the number of protons per unit energy is more than 50
times larger in the extreme starburst regions than in the molecular
disk.

\begin{figure*}[t]
\centering
\includegraphics[width=.3\textwidth,height=7cm]{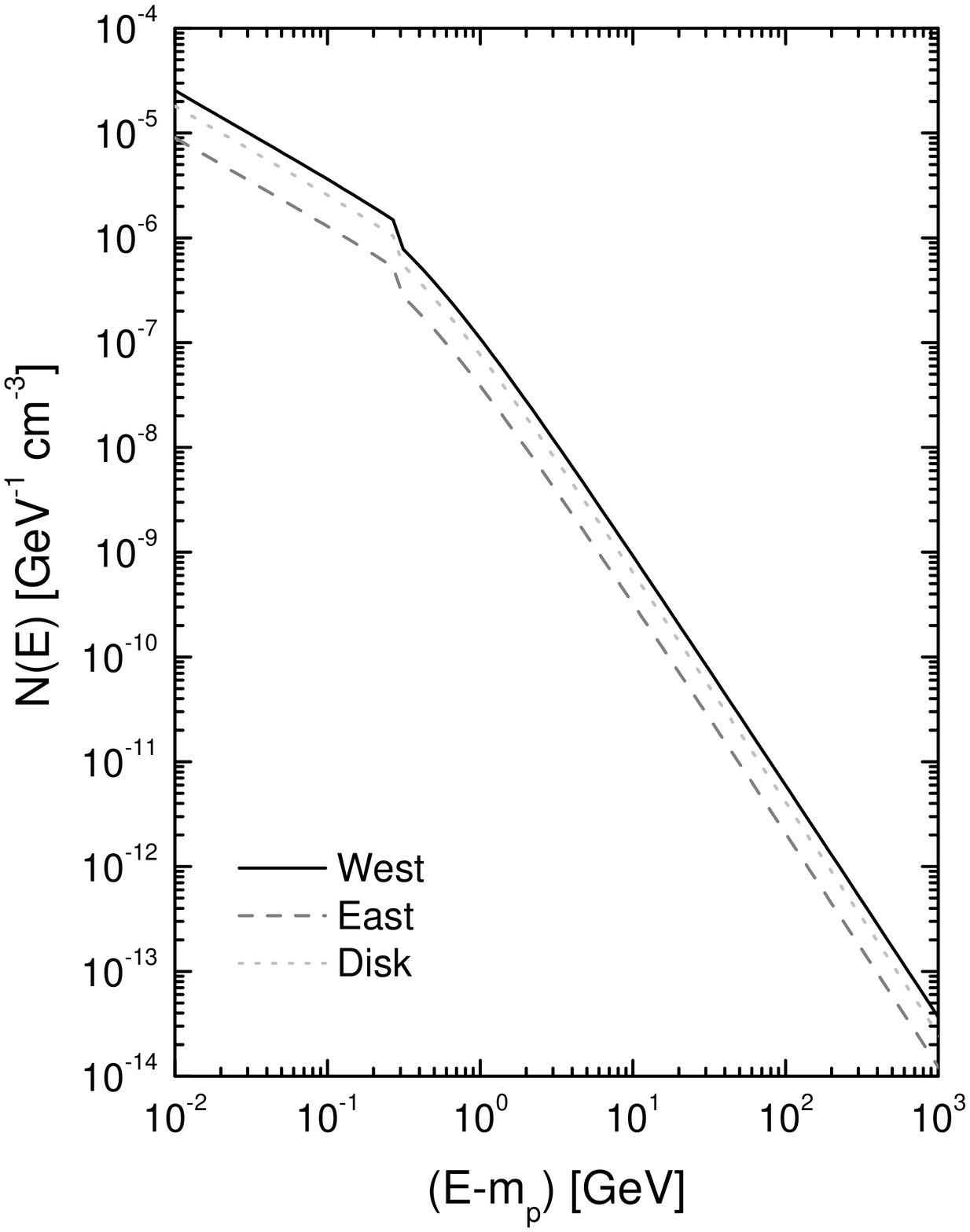}\hspace{0.5cm}
\includegraphics[width=.3\textwidth,height=7cm]{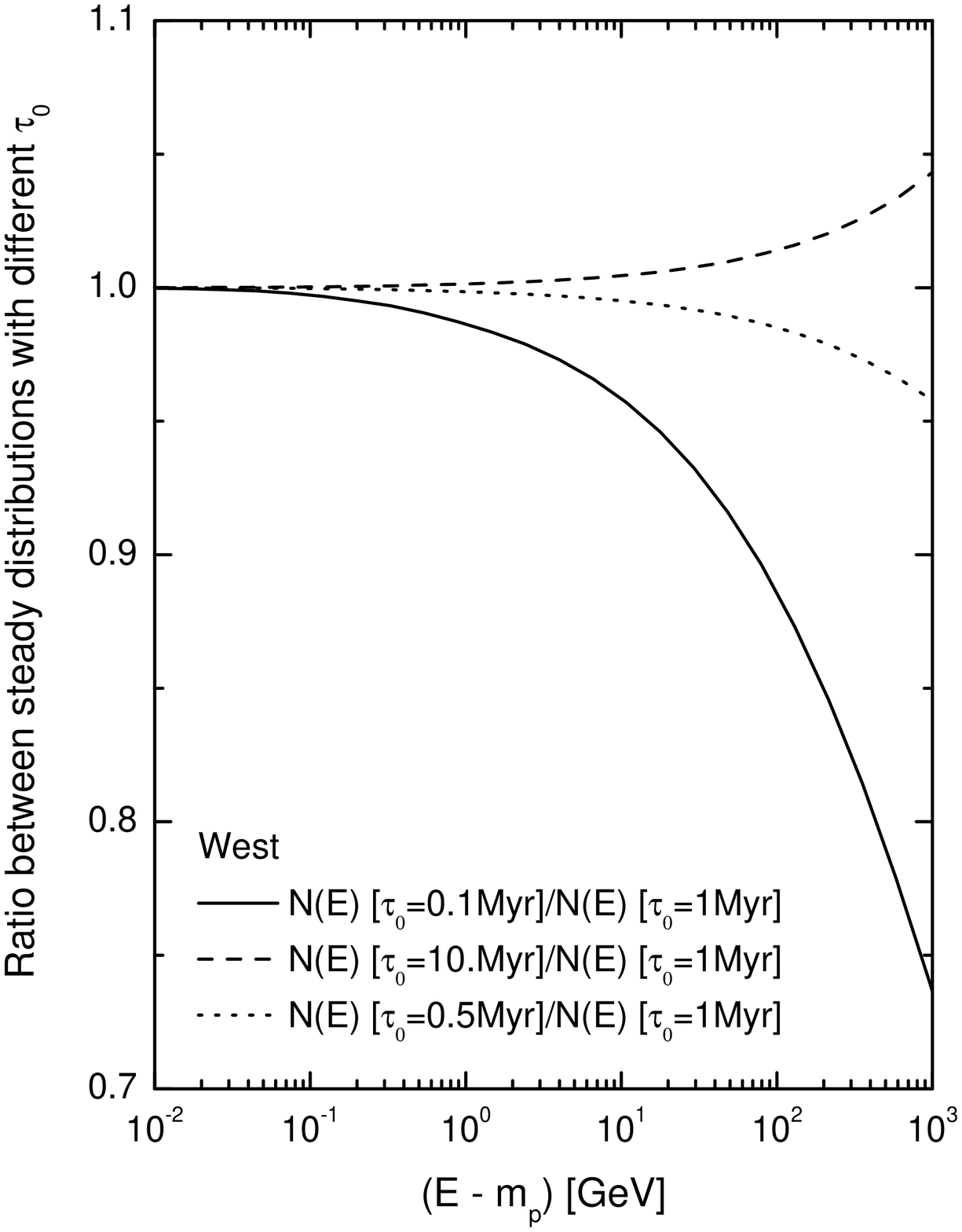}\hspace{0.5cm}
\includegraphics[width=.3\textwidth,height=7cm]{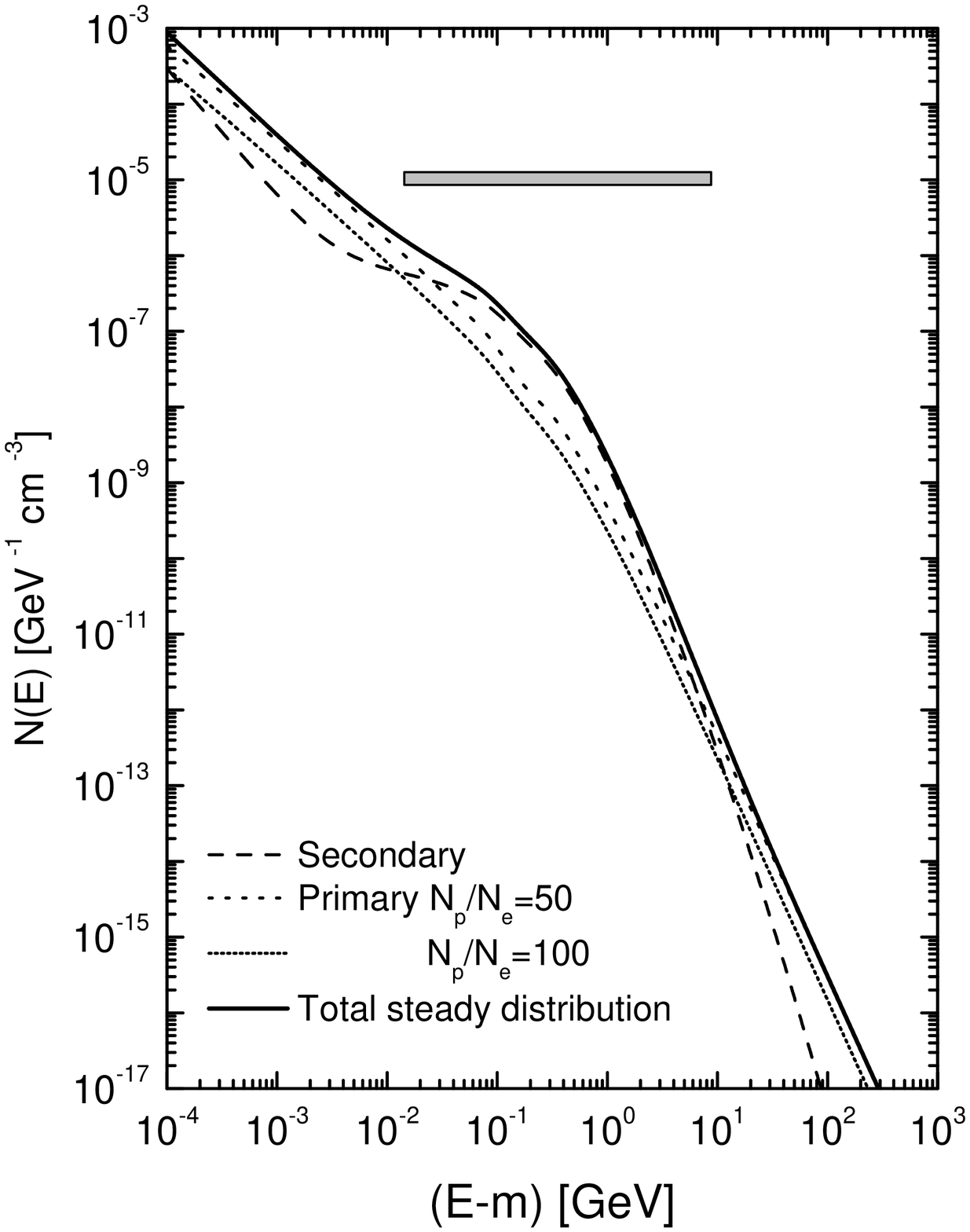}\hspace{0.5cm}
\caption{Left: Steady distribution of protons in each of the
components of Arp 220. Middle: Testing the influence of the
parameter $\tau_0$ in the determination of the proton steady
distribution. Right: Example for a steady distribution of electrons
and positrons in a western-like starburst (with $B=10$ mG). The
contribution to the total steady distribution of the primary and
secondary electrons and positrons is separately shown. The
horizontal rectangle shows the region of electron kinetic energies
where the steady distribution of secondary electrons is larger than
that of the primary electrons. It is in this region of energies
where most of the synchrotron radio emission is generated. }
\label{steadyarp}
\end{figure*}

\subsection{Electrons and positrons}

With the steady proton spectrum shown in Figure \ref{steadyarp},
left panel, the knock-on, and pion-generated electron and positron
emissivities are computed. To these emissivities, an injection
electron spectrum is also added, which is assumed as the proton
injection times a scaling factor; the inverse of the ratio between
the number of protons and electrons, $N_p/N_e$ (e.g., Bell 1978).
This ratio is about 100 for the Galaxy, but could be smaller in star
forming regions, where there are multiple acceleration sites. For
instance, V\"olk et al. (1989) obtain $N_p/N_e \sim 30$ for M82.
$N_p/N_e=100$ is assumed for the disk and $N_p/N_e=50$ is assumed
for both of the starburst nuclei. These values stand for a
conservative approach, e.g. the more the primary electrons, the
larger the inverse Compton $\gamma$-ray emission.

With such emissivities, and using the diffusion-loss equation with
corresponding losses, the leptonic steady distribution is
calculated. The inverse Compton scattering losses make use of the
photon density in the FIR derived above, and additionally, a value
of magnetic field is assumed to compute the influence of synchrotron
losses. The difference between the primary and secondary electrons
steady distributions, for a western-like extreme starburst with a
magnetic field of 10 mG is shown in the right panel of Figure
\ref{steadyarp}.

\subsection{Radio emission and magnetic fields}

\begin{figure*}[t]
\centering
\includegraphics[width=.4\textwidth]{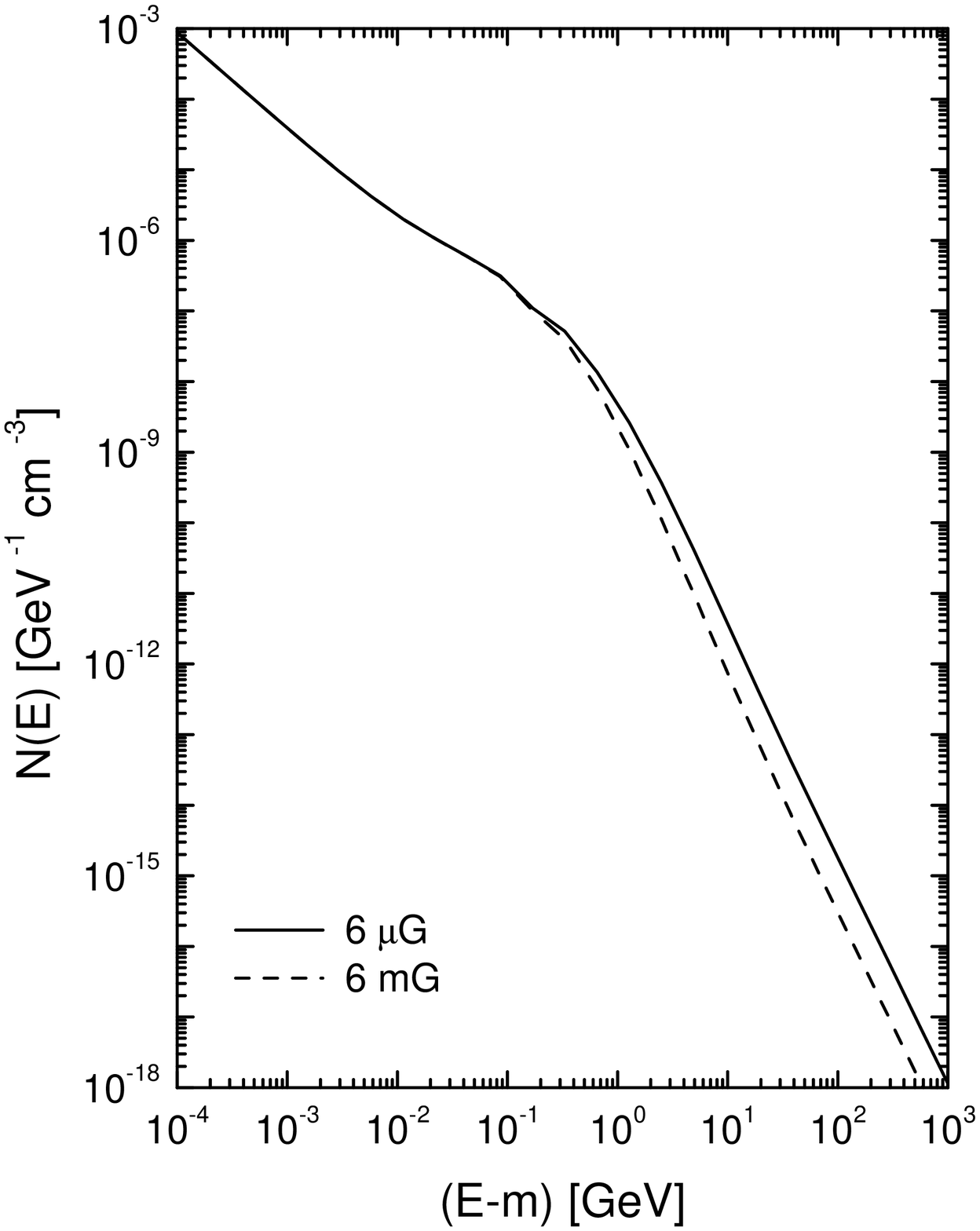}\hspace{0.5cm}
\includegraphics[width=.4\textwidth]{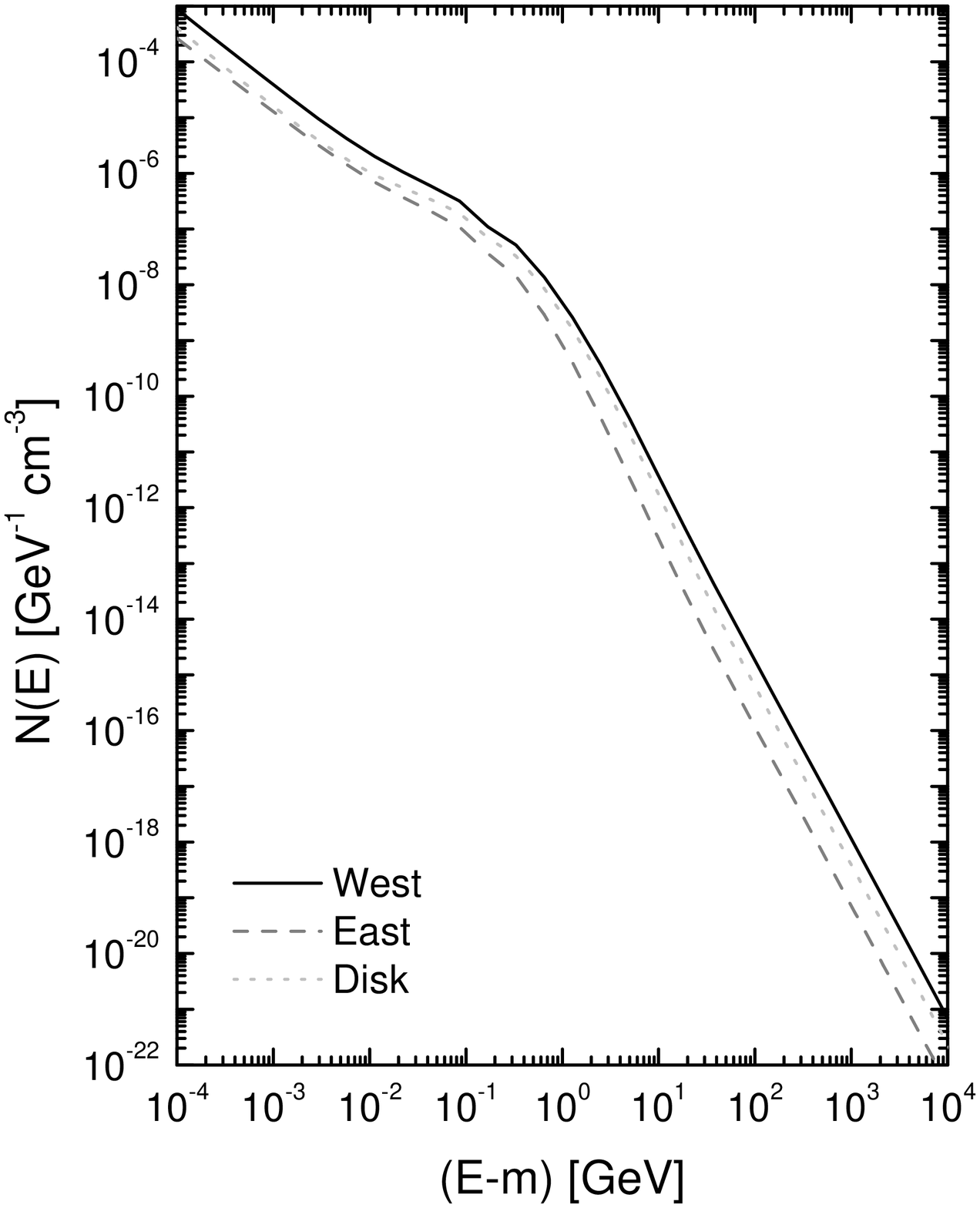}\hspace{0.5cm}
\caption{Left: Influence of the magnetic field in the determination
of the steady state electron distribution in a western-like
starburst region. Right: Final steady leptonic distributions whose
radio emission fit observations. It is with these distributions that
leptonic $\gamma$-ray emissivities are computed. } \label{b-inf}
\end{figure*}

The influence of the magnetic field upon the steady state electron
distribution is shown in Figure \ref{b-inf}. The greater the field,
the larger the synchrotron losses --what is particularly visible at
high energies, where synchrotron losses play a relevant role. Thus,
the larger the field the smaller the steady distribution. These
effects evidently compete between each other in determining the
final radio flux. In order to model the different components of Arp
220, the magnetic field is required to be such that the radio
emission generated by the steady electron distribution in each
region (see Appendix) is in agreement with the observational radio
data. This is achieved by iterating the feedback between the choice
of magnetic field, the determination of the steady distribution, and
the computation of radio flux [and at the same time taking into
account free-free emission and absorption processes, see Appendix].
These distributions are shown in the right panel of Figure
\ref{b-inf}. To reproduce the observational radio data, it is
important to note that whereas free-free emission is  subdominant
when compared with the synchrotron flux density, free-free
absorption plays a key role at low frequencies, where it determines
the opacity.

\begin{figure*}[t]
\centering
\includegraphics[width=.4\textwidth]{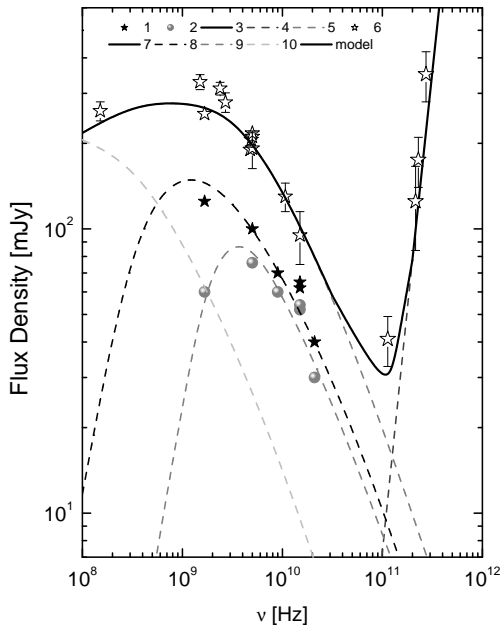}\hspace{0.5cm}
\caption{Radio and FIR emission of the different components of Arp
220. 1, 2, 3 (also compiled by Downes \& Solomon 1998, see their
figure 19), and 4 (see Figure \ref{dust}) are observational data
points corresponding to the western and eastern nuclei, the disk,
and the total FIR flux density, respectively. The curves close to
each of these set of points is the result of the modelling. The
curve with no observational data points nearby is the model
prediction for the molecular disk; its emission is summed into the
thickest black curve, which is the result of adding all components
and the final prediction of the model for the radio emission. }
\label{radiofig}
\end{figure*}

The radio emission produced by these distributions is shown in
Figure \ref{radiofig}, together with observational data. The beam
size for the different data points varies (see, e.g., table 3 of
Soop \& Alexander 1991) and unless in the cases of sub-arcsec
observations, in general, the beam contains a region larger than the
one modelled herein. However, it is expected that most of the radio
emission comes from the central and more active regions of the
galaxy, and thus a reasonable model of the nuclear environment
should reproduce most of the radiation. The magnetic field and the
free-free critical frequencies for each of the components are given
in Table \ref{radio-table}. The solid curve in Figure \ref{radiofig}
is, then, not a fit to the data, but the prediction of the
theoretical model with the chosen magnetic field. This prediction
takes into account the presence of secondary electrons, which, as
can be seen in the right panel of Figure \ref{steadyarp}, dominate
the steady distribution in the energy range where most of the radio
emission is produced. The FIR observations and modelling shown in
Figure \ref{radiofig} is that already presented in Figure
\ref{dust}: it can be noted here that the observational data point
at $\nu \sim 10^{11}$ Hz is accounted for when considering the
contribution of the non-thermal radio emission at that frequency.

The lowest frequency data point in each of the components is used to
define the critical frequency for the free-free opacity. This is a
function of the emission measure and temperature. But since there is
only one observational point at such low values of $\nu$, the
reliability of the determination of the critical frequency is lower
than that of the magnetic field. The latter is the main responsible
for the fixing of the steady electron distribution and the
prediction of the radio emission at higher frequencies, where
several observations are available for comparison.

To exemplify the uncertainty in the critical frequency
determination, consider the western nucleus. In that case, the
lowest frequency point could be thought of as being part of the
free-free opacity-produced decay of the radio emission curve, or as
part the non-thermal synchrotron trend, if the critical frequency is
lower. An intermediate situation is adopted here. This also
influences the value of critical frequency adopted for the disk
--forcing the critical frequency in that case to be lower than that
in the extreme starbursts in order to be in agreement with the first
data point of the total radio curve. For the eastern nucleus, it is
apparently clear that the first data point --obtained at high
angular resolution-- is already opacity-dominated, since its value
is less than the contiguous data at higher frequency. In any case,
both nuclei seem to have a relatively high critical frequency,
particularly when compared with the disk, what would be in agreement
with them being stronger star forming regions.\footnote{In passing,
note also that the turnover of the spectrum happens at too high a
frequency as to be produced by synchrotron self-absorption, e.g. by
using the sizes of Arp 220 components, and Eq. 3.56 of Kembhavi and
Narlikar (2001).} The critical frequencies mentioned in Table
\ref{radio-table} can be obtained with temperatures between 5 and 10
$\times 10^3$ K, and EM values between 10$^4$ and 10$^7$ pc
cm$^{-6}$, the smaller EM corresponding to the disk. Similar values
of critical frequencies, temperatures, and emission measures were
used to model the radio emission in the case of the starburst galaxy
NGC 253 (Paglione et al. 1996).


\begin{deluxetable}{lcc}
\tablewidth{0.65\textwidth}
\tablecaption{Parameters for radio modelling.}

\tablehead{ \colhead{Component} & \colhead{Magnetic Field } &
\colhead{Critical Frequency }

} \startdata

western starburst & 6.5 mG & 0.38 GHz \nl 

eastern starburst & 4.5 mG & 2.86 GHz  \nl 

disk & 280 $\mu$G & 0.07 GHz \nl 


\enddata
\label{radio-table}
\end{deluxetable}

Consider now the analysis of the magnetic field results, which
appear, as said, to be more stable against model degeneracy. It is
worth noticing that not much is known about the magnetic field in
ULIRGs, except for upper limits ($\sim 5$ mG), obtained with Zeeman
splitting measurements of four southern OH megamaser galaxies
(Killeen et al. 1996). This study, being for a more active
star-forming galaxy, is compatible with these estimates and favor
the ideas regarding the existence of such high fields in extreme
starbursts (e.g., Smith et al. 1998).

It is to be remarked that for both the western and eastern nuclei,
the minimal energy argument does not seem to hold.\footnote{The
magnetic field strength in a galaxy produces an energy density that
can be compared with the energy density stored in the relativistic
populations of particles. When these densities are similar, the
system is said to be under energy equipartition (see, e.g., Kembhavi
\& Narlikar 2001, p. 50).} With the magnetic field strength given in
Table \ref{radio-table}, and the relativistic steady state
populations of Figures \ref{steadyarp} (left panel) and \ref{b-inf}
(right panel), only the molecular disk is in magnetic energy
equipartition. This appears to be a similar scenario -although more
extreme- to that found for the interacting galaxy NGC 2276, where
the magnetic field seems not in energy equipartition with cosmic
rays either (Hummel \& Beck 1995).

The magnetic field in the extreme starbursts is compatible with
those measured nearby supernova remnants in the Galaxy (Koralesky et
al. 1998; Brogan et al. 2000), where field strengths between 0.1 and
4 mG were found. These fields strengths were interpreted as being an
ambient magnetic field compressed by the supernova remnant. The same
mechanism could be thought of for Arp 220's western and eastern
nucleus. The disk magnetic field, in turn, is compatible with the
result for molecular clouds presented by Crutcher (1991), what is in
agreement with the disk itself being thought of as a gigantic
molecular cloud with the gas filling all the medium.

Similarly high values of magnetic fields ($B > 800\, \mu$G) were
necessary to produce the observed collimated outflows in ULIRGs, and
particularly in Arp 220, as a resultant of a strong starburst
environment (Gouveia Dal Pino \& Medina Tanco 1999). Finally, the
overall magnetic field distribution bears some resemblance with our
own Galactic Center. There, in a few dense gas clouds about 2 pc
north of the Galactic center, field strengths in the milligauss
range were derived from Zeeman measurements (see Beck 2001 for a
review; Plante et al. 1994; Yusef-Zadeh et al. 1996). The average
field in Sgr A complex is, in analogy with the disk value,
restricted to less than 0.4 mG (Reich 1994). The non-detection of
the Zeeman effect in the OH lines (Uchida and G\"usten 1995) also
indicates a relatively weak general magnetic field into which clouds
with strong fields are embedded.

\subsection{$\gamma$-ray emissivity and first estimation of fluxes}

\begin{figure*}[t]
\centering
\includegraphics[width=.4\textwidth]{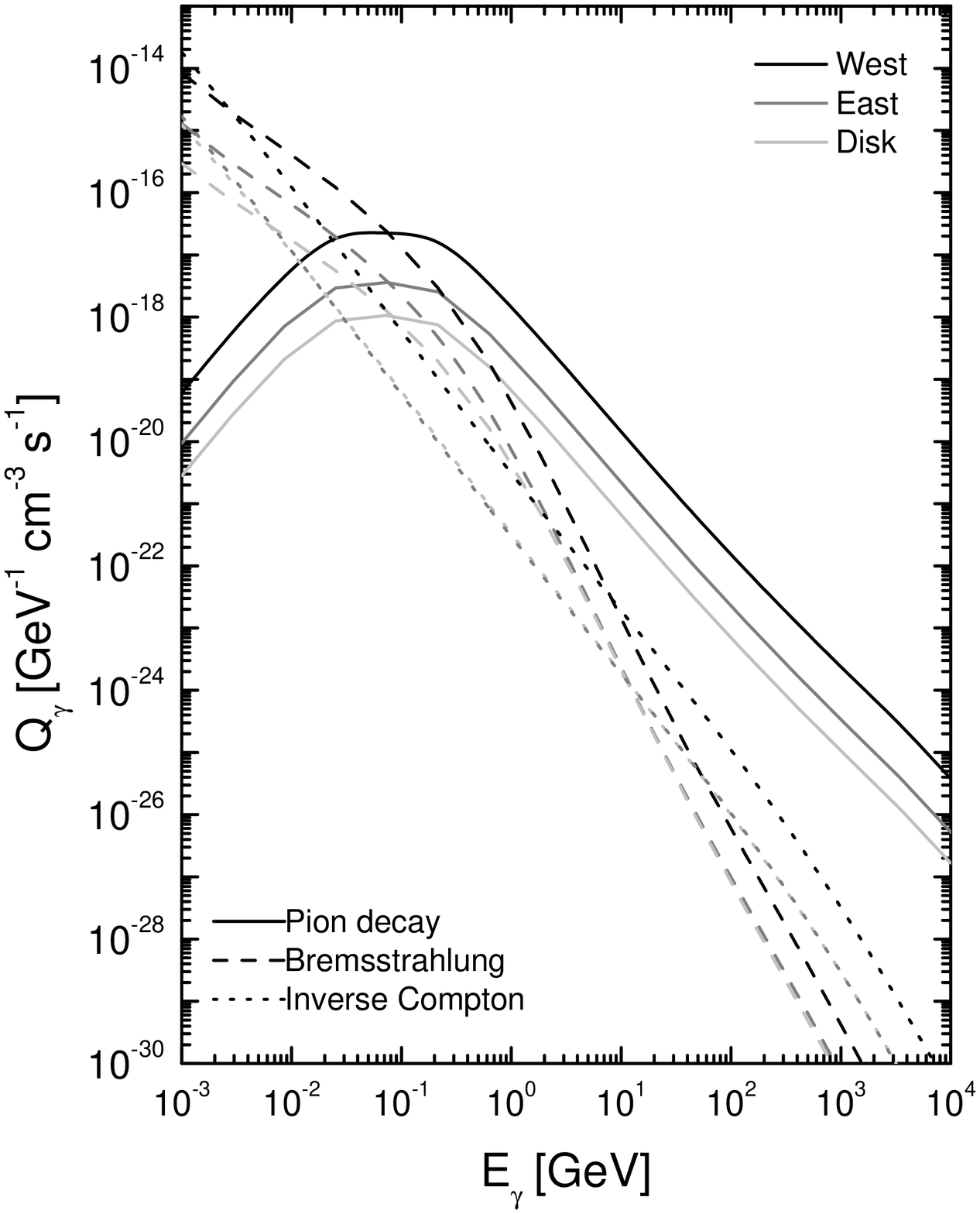}\hspace{0.5cm}
\includegraphics[width=.4\textwidth]{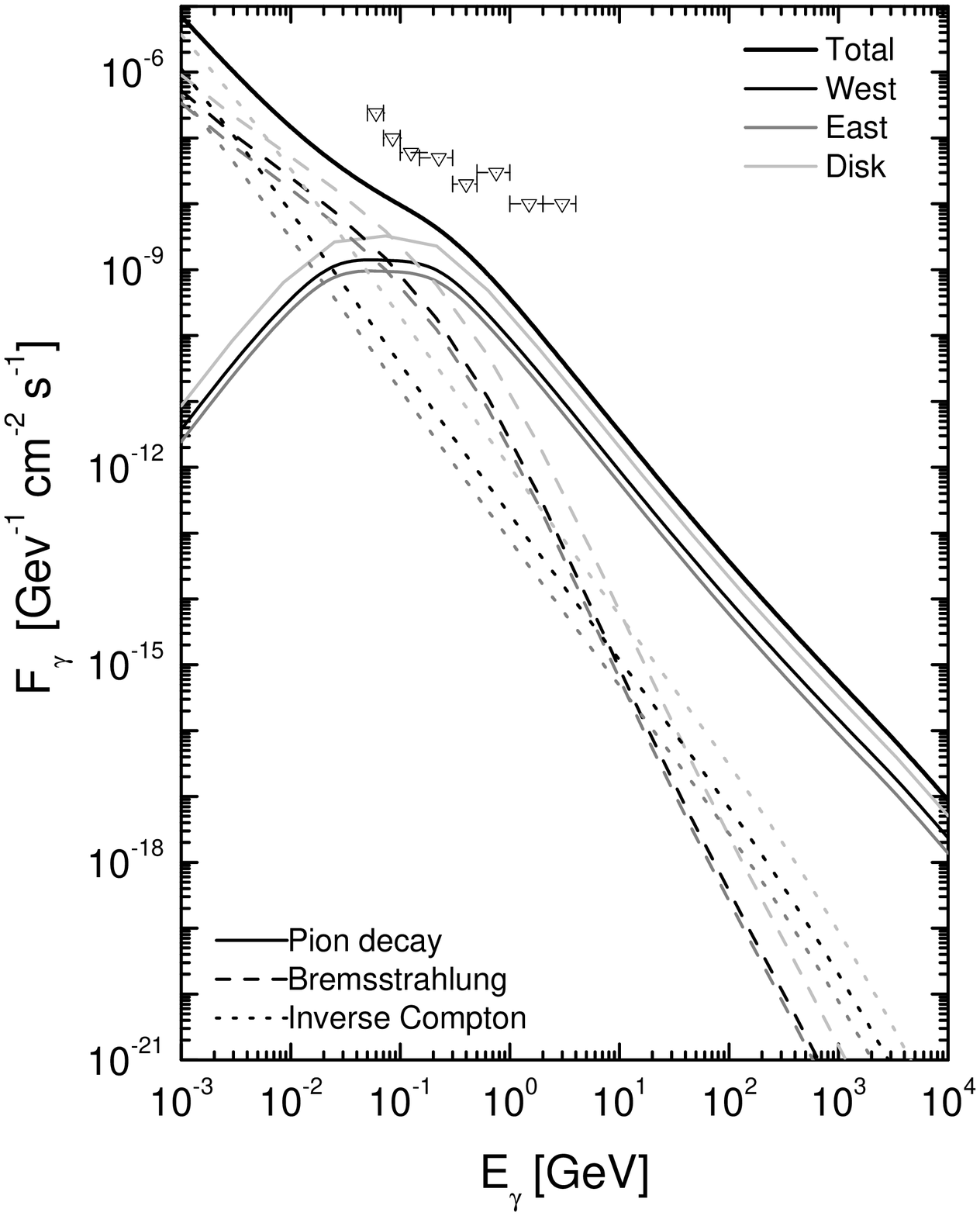}
\caption{Left: Bremsstrahlung, inverse Compton, and pion decay
emissivity of $\gamma$-rays in the different components of Arp 220.
Right: Differential fluxes without considering opacity effects. The
down-triangles are EGRET upper limits. } \label{g-emis}
\end{figure*}

\begin{figure*}[t]
\centering
\includegraphics[width=.4\textwidth]{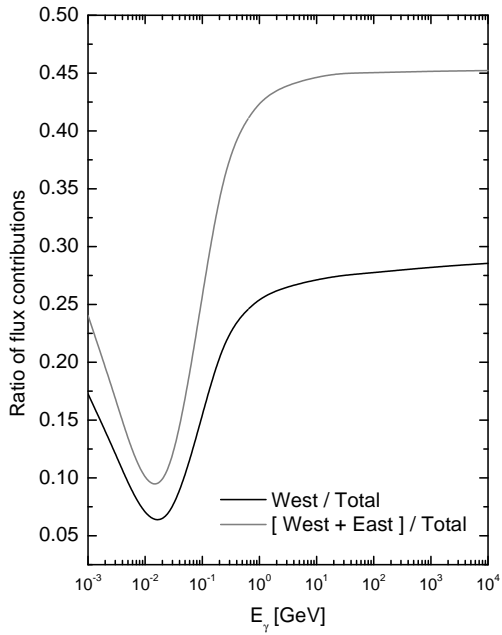}\hspace{0.5cm}
\caption{Relative importance of the extreme starburst regions as
compared with the total $\gamma$-ray flux predicted from Arp 220. No
opacities are herein considered although its inclusion would not
change this result. } \label{ratio}
\end{figure*}

In the left panel of Figure \ref{g-emis} the bremsstrahlung, inverse
Compton, and pion decay $\gamma$-ray emissivities of the different
components of Arp 220, is shown. These results are derived for the
model which is in agreement with radio and IR-FIR observations. At
energies above 100 MeV, pion decay $\gamma$-rays is the dominant
contribution, as expected. Clearly, the emissivity of high energy
photons is the largest in the western extreme starburst, the most
active region of star formation. It is followed by the eastern
nuclei, and in a subdominant role, by the molecular disk. The
differential flux, shown in the right panel of Figure \ref{g-emis}
without considering absorption effects, shows the effect of volume.
The disk $\gamma$-ray flux is the largest, and the nuclei are now
subdominant. Nevertheless, only the western starburst provides more
than one fourth of the total $\gamma$-ray flux (similar to the
weight of its contribution in the IR band; although note, however
that the total luminosity in the $\gamma$-ray band is much less than
in the IR).
%
%
The relative importance of the western and eastern nuclei in the
total $\gamma$-ray radiation budget is shown in Figure \ref{ratio}.
Upper limits to the differential photon flux from Arp 220 are also
shown in Figure \ref{g-emis}. These limits were obtained from an
analysis of 4 years of EGRET data (see Cillis et al. 2004) and are
in agreement with model predictions.


\subsection{$\gamma$-ray escape }

The opacity to $\gamma\gamma$ pair production with the photon field
which, at the same time, is target for inverse Compton processes can
be computed as $ \tau(R_{c},E_\gamma)^{\gamma\gamma}=\int
\int_{R_c}^\infty n(\epsilon)
\sigma_{e^-e^+}(\epsilon,E_\gamma)^{\gamma\gamma}  dr \, d\epsilon ,
\label{op1} $ where $\epsilon$ is the energy of the target photons,
$E_\gamma$ is the energy of the $\gamma$-ray in consideration, $R_c$
is the place where the $\gamma$-ray photon was created within the
system, and $ \sigma_{e^-e^+}(\epsilon,E_\gamma)^{\gamma\gamma} =
({3\sigma_T}/{16}) (1-\beta^2) (2\beta
(\beta^2-2)+(3-\beta^4)\ln((1+\beta)/(1-\beta))), $ with
$\beta=(1-(m c^2)^2/(\epsilon \, E_\gamma))^{1/2}$ and $\sigma_T$
being the Thomson cross section, is the cross section for
$\gamma\gamma$ pair production (e.g. Cox 1999, p.214). Note that the
lower limit of the integral on $\epsilon$ in the expression for the
opacity is determined from the condition that the center of mass
energy of the two colliding photons should be such that $\beta
>0$. The fact that the dust within the starburst
reprocesses the UV star radiation to the less energetic infrared
photons implies that the opacities to $\gamma\gamma$ process is
significant only at the highest energies.
It can be seen that $\tau(R_{c},E_\gamma)^{\gamma\gamma} <
\tau(E_\gamma)_{\rm max}^{\gamma\gamma} = 2R \int_0^\infty
n(\epsilon) \sigma_{e^-e^+}(\epsilon,E_\gamma) d\epsilon \, $, since
no source of opacity outside the system under consideration is
assumed, whose maximum linear size in the direction to the observer
is, in the case of a sphere of radius $R$, equal to $2R$. For the
molecular disk, $ \tau(E_\gamma)_{\rm max}^{\gamma\gamma}=(h/\cos
i)\int_0^\infty n(\epsilon) \sigma_{e^-e^+}(\epsilon,E_\gamma)
d\epsilon \, $.

The opacity to pair production from the interaction of a
$\gamma$-ray photon in the presence of a nucleus of charge $Z$ needs
to be considered too. Its cross section in the completely screened
regime ($E_\gamma / mc^2 \gg 1/(\alpha Z)$) is independent of
energy, and is given by (e.g. Cox 1999, p.213) $
\sigma_{e^-e^+}^{\gamma Z} = (3 \alpha Z^2 \sigma_T / 2 \pi ) (7/9
\ln (183/Z^{1/3}) - 1/54)$. At lower energies the relevant cross
section is that of the no-screening case, which is logarithmically
dependent on energy, $ \sigma_{e^-e^+}^{\gamma Z} = (3 \alpha Z^2
\sigma_T / 2 \pi ) (7/9 \ln (2E_\gamma/mc^2) - 109/54)$, and matches
the complete screening cross section at around 0.5 GeV. Both of
these expression are used to compute the opacity, depending on
$E_\gamma$. Use of the fact that the cross section, in typical ISM
mixtures of H and He, is $\sim 1.3$ times bigger than that of H with
the same concentration, is also made and the opacity is accordingly
increased (see, e.g., Ginzburg \& Syrovatskii 1964, p. 30).

From the properties deduced from the radio emission, i.e. the
magnetic field and emission measure in each of Arp 220 components,
it can be seen that Compton scattering and attenuation in the
magnetic field by one-photon pair production are negligible.

\begin{figure*}[t]
\centering
\includegraphics[width=.4\textwidth]{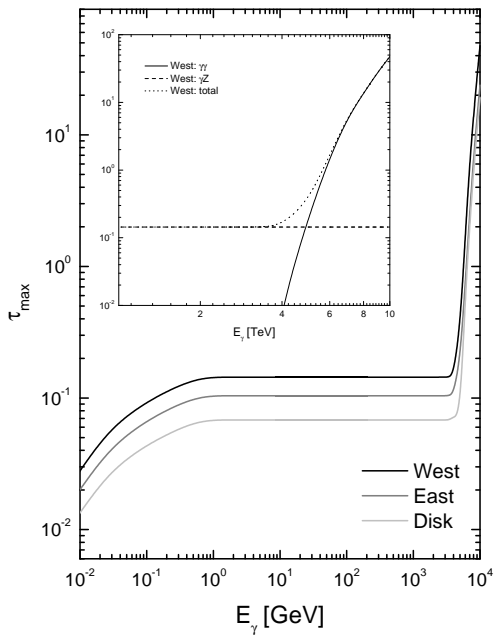}\hspace{0.5cm}
\caption{Opacities to $\gamma$-ray scape in the different components
of Arp 220 as a function of energy. The highest energy is dominated
by $\gamma\gamma$ processes, whereas $\gamma Z$ dominates the
opacity at low energies. Significant $\tau_{\rm max}$ are only
encountered above 1 TeV, the inset shows the total, and the
contributions to the total opacity, in the case of the western
nucleus of Arp 220 for this range of energy.} \label{opaarp}
\end{figure*}

\begin{figure*}[t]
\centering
\includegraphics[width=.45\textwidth,height=8cm]{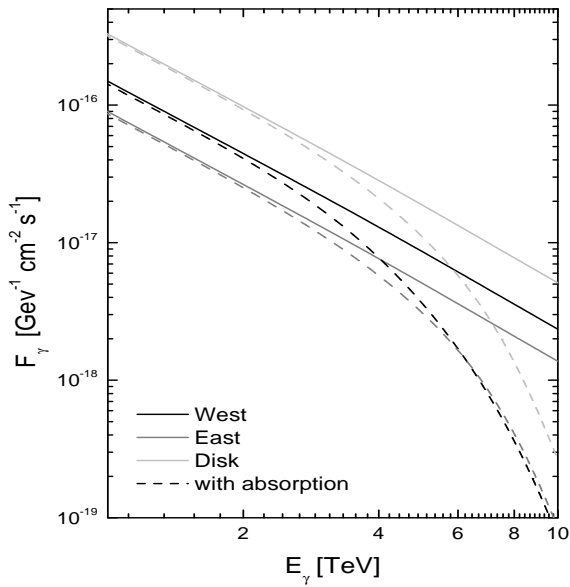}\hspace{0.5cm}
\caption{Fluxes with and without absorption processes being
considered. Appreciable differences appear only at the highest
energies. } \label{tev-abs}
\end{figure*}

In Figure \ref{opaarp}, both, the different contributions to the
opacity from $\gamma\gamma$ and $\gamma Z$, in the case of the
western starburst, and the total opacity for the three Arp 220
components are shown. The western nucleus is subject to the biggest
opacities, its value is $\sim 0.1$ up to $\sim 4$ TeV and then
rapidly increases. The equation of radiation transport (see
Appendix), for the molecular disk and extreme starburst regions, are
then used to compute the predicted $\gamma$-ray flux taking into
account all absorption processes. The smallness of $\tau_{\rm max}$
throughout most of the energy range implies that the correction
factors to the fluxes are only a few percent up to TeV energies (it
is not possible to see the difference in a plot like that presented
in the right panel of Figure \ref{g-emis}). In Figure \ref{tev-abs}
the effect of TeV photon absorption in each of the components of Arp
220 is shown in detail. Note that the disk is subject to relatively
lower opacities than the eastern and western extreme starbursts.
This is caused mainly by a reduction of the photon target density
(i.e. a reduction in $\tau^{\gamma\gamma}$ when compared with the
corresponding values found in the extreme star forming regions).


\subsection{Observability}

The total predicted flux in $\gamma$-rays above 100 MeV,  after the
effects of absorption are taken into account at all energies, is
$2.8 \times 10^{-9}$ photons cm$^{-2}$ s$^{-1}$. This is comfortably
below the upper limit for this galaxy imposed with EGRET data by
Torres et al. (2004) in the same energy range, which is about one
order of magnitude larger. It is, however, above the threshold for
detection with GLAST: $ F(>100\; {\rm MeV}) \sim 2.4 \times 10^{-9}
\;{\rm photons\; cm^{-2} s^{-1}} $ is the GLAST satellite
sensitivity for a 5$\sigma$ detection of a point-like, high latitude
source after 1 yr of all-sky survey. If this model bears resemblance
with reality, then, it might be possible for GLAST to detect  Arp
220 for the first time in $\gamma$-rays.

By the same token, the total predicted fluxes in $\gamma$-rays above
300 GeV and 1 TeV are $\sim 2 \times 10^{-12}$ photons cm$^{-2}$
s$^{-1}$ and $\sim 7 \times 10^{-13}$ photons cm$^{-2}$ s$^{-1}$,
respectively. These fluxes are high enough as to render possible,
again in the case this model bears resemblance with reality, to
detect Arp 220 at higher energies.
Reliability of the flux predictions above 1 TeV also depends on
the cross section modelling being reasonably
correct.\footnote{{\bf NOTE ADDED: This paper have used the
parameterizations of the cross section for neutral pion production
from Blattnig et al. (2000). This have later been shown to
overestimate the gamma-ray fluxes at energies above 100 GeV (see
Appendix of Domingo-Santamar\'{\i}a \& Torres 2005 for full
details). Reanalysis with different cross section
parameterizations shows that more than 100 hours are needed in
IACTs to detect the galaxy within this model. The multifrequency
modelling and results other than gamma-ray yield at the highest
energies are unaffected.}}

\v{C}herenkov telescopes cannot typically observe at zenith angles
much larger than 70$^\circ$. The zenith angle $\vartheta$ at the
upper culmination of an astronomical object depends on the latitude
$\phi$ of the observatory and the declination DEC of the object
according to $ \vartheta = | \phi - \mathrm{DEC} | $. Therefore, the
condition $ | \phi - \mathrm{DEC} | \leq 70^o$ has to be imposed in
the selection of observable objects. For the next generation (but
already operating) \v{C}herenkov telescopes and because of location,
Arp 220 seems to be a good candidate for a northern hemisphere
observatory [e.g. MAGIC has $ \vartheta \sim 5.5^o$; VERITAS has $
\vartheta \sim 9^o$]. However, it seems also possible (see Petry
2001) for HESS to observe Arp 220 at high zenith angles, since
DEC$_{\rm Arp 220}<+37^o$ implying $ \vartheta < 60^o$.


As a function of $ \vartheta$, an increase in effective collection
area is accompanied by a proportional increase in hadronic
background rate, such that the gain in flux sensitivity is therefore
only the square-root of the gain in area (Petry 2001). In addition,
the higher the value of $ \vartheta $, the higher is the energy
threshold for observation, what reduces the integral flux. If
$F_{5\sigma}( E>E_{\mathrm{thr}})$ is defined as the integral flux
above the energy threshold $E_{\mathrm{thr}}$ which results in a $5
\sigma$ detection after 50~h of observation time, $ F_{5\sigma}(
E>E_{\mathrm{thr}}(\vartheta), \vartheta ) =
      F_{5\sigma}( E>E_{\mathrm{thr}}, 0^\circ ) \cdot
\cos(\vartheta). $ The needed observation time to observe a source
with flux $F_{5\sigma}( E>E_{\mathrm{thr}})$ can be conservatively
estimated as (Petry 2001) $  T_{5\sigma}(E>E_{\mathrm{thr}}) =
\left( {F(E>E_{\mathrm{thr}})}/{F_{5\sigma}(E>E_{\mathrm{thr}})}
\right)^{-2}  50 \, \mathrm{hours}. $ In the case of the modelling
herein presented for Arp 220, assuming a generic, but conservative,
$F_{5\sigma}( E>E_{300 {\rm GeV}}, 0^\circ )=3 \times 10^{-12}$
photons cm$^{-2}$, the needed observation time for the galaxy to
appear above 300 GeV is about 95 hours.

Finally, note that the decay of charged pions will also lead to the
production of energetic neutrinos. While the analysis of the
neutrino production and possible observability of Arp 220 by the
future neutrino telescopes is left to a subsequent publication, we
note that the flux of neutrinos that is outcome of this model would
not violate the upper limits imposed by the AMANDA II experiment
(Ahrens et al. 2004). Even if the neutrino flux from Arp 220, is the
same as the photon flux, it would be
below imposed upper limits to the fluxes from all candidate neutrino
sources.

\section{Concluding remarks}

\begin{figure*}[t]
\centering
\includegraphics[width=.35\textwidth,height=8cm]{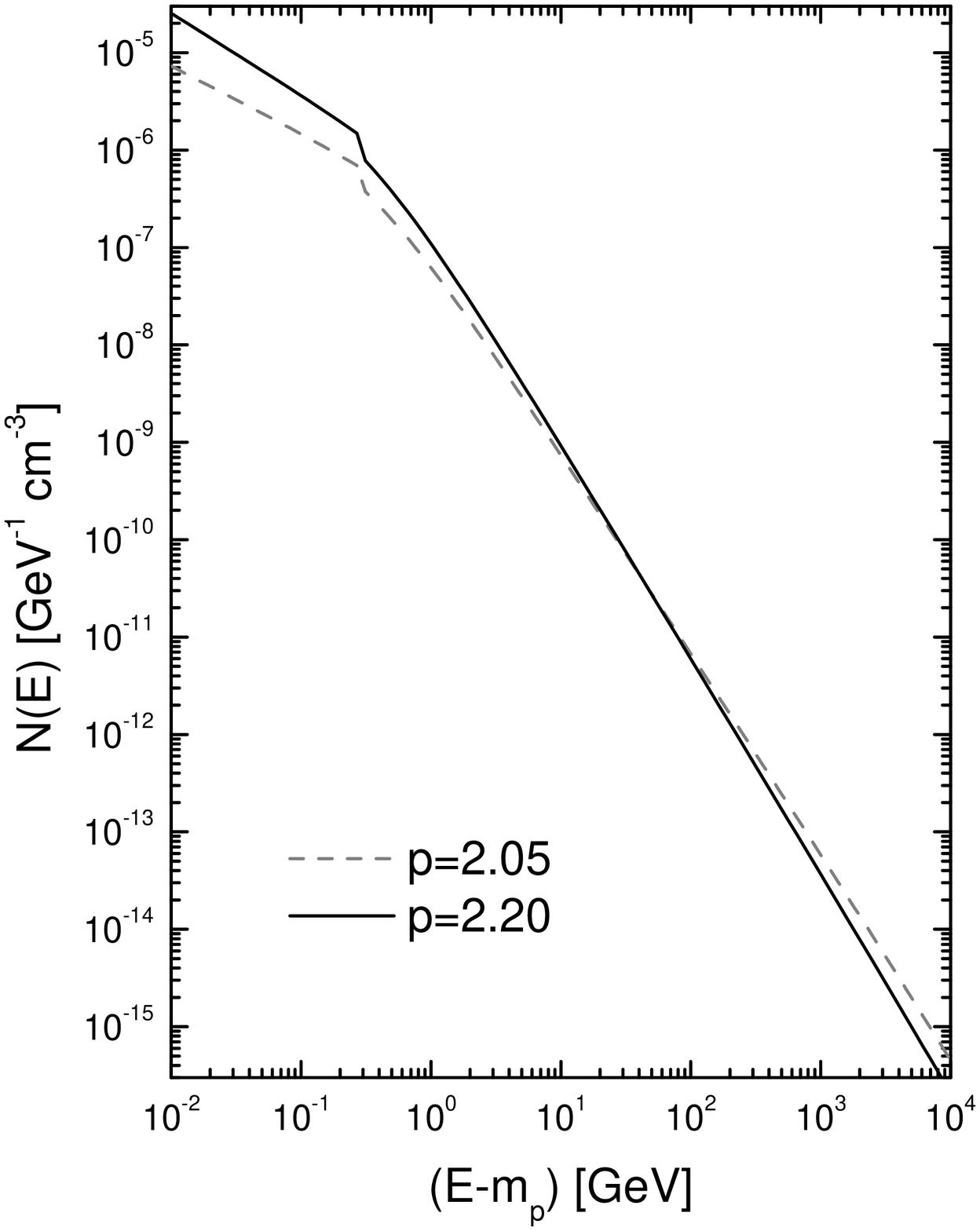}\hspace{0.5cm}
\includegraphics[width=.35\textwidth,height=8cm]{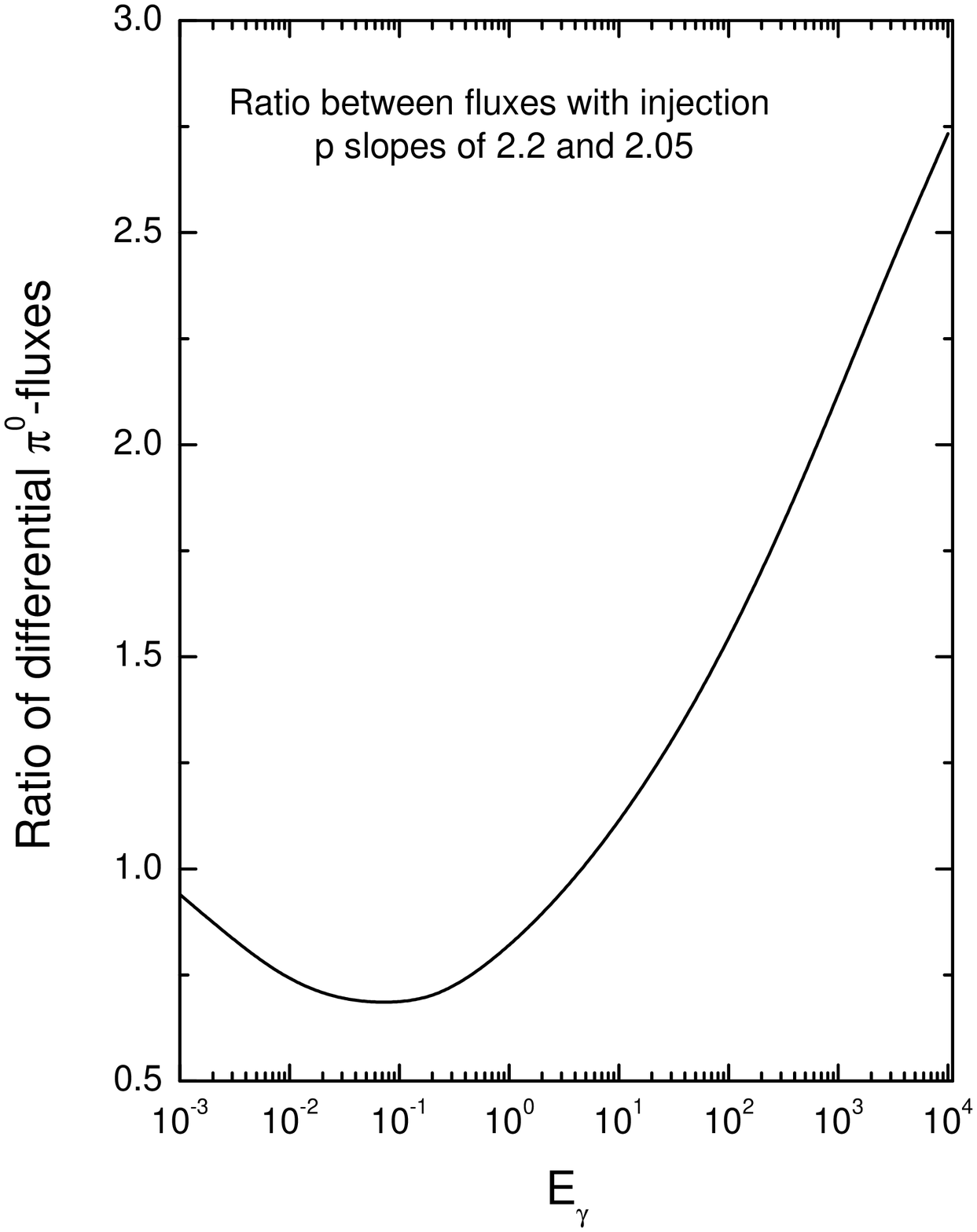}
\caption{Left: dependence of the steady state proton distribution on
the proton injection power law slope, $p$. Right: ratio between pion
decay fluxes, in the western nuclei of Arp 220, for different proton
injection power law slopes. } \label{slope}
\end{figure*}

Luminous infrared galaxies are certainly interesting objects, and
until recently, focus on them have been mainly granted at all
wavelengths but one, the high energy domain. With several new
\v{C}erenkov telescopes, $\gamma$-ray satellites, cosmic ray, and
neutrino observatories on the verge of becoming operational, or
operating already, the interest on the possible high energy features
of LIRGs and ULIRGs has been rekindled. There is much to learn at
high energies, whether these galaxies are detected or not.
Sensitivities of forthcoming equipments is --as discussed above--
high enough as to impose severe constraints on theoretical models or
provide interesting clues in our understanding of these objects.

Recently, ULIRGs have been analyzed as possible ultra high energy
cosmic ray sources (Smialkowski, Giller \& Michalak 2002; Torres \&
Anchordoqui 2004), and yet unidentified $\gamma$-ray detections
(Torres et al. 2004; Torres 2004; Cillis et al. 2004). In this
paper, a self-consistent model for the radio, IR, and $\gamma$-ray
emission from Arp 220, the prototypical and nearest ULIRG, was
presented. Complete agreement with observational data was obtained
at all frequencies, and predictions of $\gamma$-ray fluxes were
obtained. These fluxes suggest that Arp 220 could be a source for
GLAST as well the new \v{C}erenkov telescopes. The radio emission
modelling of Arp 220, as the result of primary and secondary
electrons' synchrotron emission, appear to indicate that the central
regions of Arp 220 are subject to a strong magnetic field.

Although many are the free parameters involved in this modelling,
few are those which are unrelated to observations, and even fewer
are those which --if changing-- may have a significant impact on the
results. Consider, as an example, the choice of the power slope for
the injected proton spectrum. The model presented assumed it to be
2.2 (i.e. $Q_{\rm inj} \propto E^{-2.2}$), for all three Arp 220
components analyzed. However, there is nothing a priori yielding to
this value, except that it is a reasonable and conservative
expectation for the slope of a relativistic proton population in the
vicinity of its acceleration site, e.g. a supernova remnant shock.
But perhaps, given that the western extreme starburst is the
strongest site of star formation known, the proton population might
have there a harder spectrum, in particular, as compared with that
found in the disk. Figure \ref{slope} explores how a change in the
injected proton spectrum would affect the results. The left panel
compares the steady state proton distribution for a 2.2 (the
previously assumed slope) and a 2.05 spectrum. Since the same power
is injected with a harder slope, the latter spectrum dominates at
high energies. The right panel shows that the ratio between --for
instance-- $\gamma$-ray fluxes produced in pion decays in the
western nucleus would not change much as a function of energy,
although in the direction of favoring the possible detection.

It is also interesting to note that the electron steady
distribution, interacting via inverse Compton with the abundant IR
photons, will also contribute to the flux at lower frequencies, i.e.
in the hard X-ray regime. Thus, a diffuse model for the high energy
emission also needs to yield fluxes in agreement with imposed upper
limits at  hard X-ray/soft $\gamma$-ray frequencies.  Dermer et al.
(1997) found using OSSE that the photon flux is less than $1.2
\times 10^{-4}$ photons cm$^{-2}$ s$^{-1}$, and $0.7 \times 10^{-4}$
photons cm$^{-2}$ s$^{-1}$, in the 0.05--0.10 and 0.10--0.20 MeV
bands, respectively. The luminosity limit in the whole energy range
mentioned is $3 \times 10^{43}$ erg s$^{-1}$. Iwasawa et al. (1999)
found using Beppo-Sax a luminosity upper limit of $5 \times 10^{40}$
and $1 \times 10^{41}$ erg s$^{-1}$ in the 0.5--2 and 2--10 keV
bands, respectively. These are also consistent with previously
imposed ASCA limits, and stringent than those limits imposed using
Chandra at such hard X-ray energies. The model discussed in this
work yields inverse Compton fluxes of a few percent or less than the
mentioned upper limits at these energies. Moran et al. (1999) found,
although with a less detailed modelling, a similar situation in the
galaxy NGC 3256. This is consistent with the hard X-ray/soft
$\gamma$-ray emission being mostly generated not by diffuse
processes, but by several powerful point sources, which is also the
case, according to recent INTEGRAL observations (Lebrun et al.
2004), in our Galactic Center.

In closing, three remarks are deemed important to keep in mind when
analyzing possible observations of Arp 220 and other LIRGs a)
Additional hadronic production of high energy $\gamma$-rays with
matter in the winds of stars (Romero \& Torres 2003; Torres et al.
2003), and emission from particular stellar systems in general
(e.g., Romero et al. 1999, Benaglia et al. 2001, Benaglia \& Romero
2003) was herein disregarded, and although subdominant, it would
certainly help in increasing the $\gamma$-ray emission at the
highest energies. b) Only non-variable $\gamma$-ray sources can be
ascribed to LIGs if diffusive process such as the one explored in
this work are responsible for the emission. Variability indices
(Torres et al. 2001, Nolan et al. 2003) could then help in
discriminating unidentified detections. c) The small redshift of Arp
220 and other galaxies in the 100 Mpc sphere makes opacities due to
processes with photons of the cosmic microwave and IR background
outside the galaxy negligible below 10 TeV (see, e.g.,
figure 2 of Aharonian 2001).

\subsection*{Acknowledgements}

This work was performed under the auspices of the U.S. D.O.E. (NNSA)
by the University of California Lawrence Livermore National
Laboratory, under contract No. W-7405-Eng-48. I thank Felix
Aharonian, Luis Anchordoqui, Steve Blattnig, Valenti Bosch-Ram\'on,
Analia Cillis, Thomas Dame, Seth Digel, Eva Domingo-Santamar\'{\i}a,
Yu Gao, Felix Mirabel, Martin Pessah, Olaf Reimer, Gustavo Romero,
Nick Scoville, \& Wil Van Breugel for comments related to this work.
Eva Domingo, Seth Digel, and Analia Cillis are further thanked for
discussions on TeV analysis of data, GLAST sensitivity, and EGRET
upper limits, respectively. Martin Pessah is further acknowledged
for discussions regarding geometry. An anonymous referee is also
acknowledged. I finally thank Ileana Andruchow for support.

\clearpage

\section*{Appendix}

Here we present some of the main formulae used in the paper and a
few details of implementation. For a more complete account see the
SPIRES-HEP version (astro-ph/0407240) of this article.

\noindent {\bf Proton losses \label{PLOS}}

During the motion of a proton through a neutral medium, the
ionization loss rate is given by (e.g., Ginzburg \& Syrovatskii
1964, p.120ff) \be -\left( \frac{dE}{dt} \right)_{{\rm Ion},p} \sim
1.83 \times 10^{-17} \left( \frac {n_{\rm H}+2 n_{{{\rm H}}_2}}{{\rm
cm}^{-3}}\right) \frac cv \left\{ 10.9 + 2 \ln \left(\frac {E}{m_p
c^2} \right) + \ln \left( \frac {v^2}{c^2} \right) - \frac
{v^2}{c^2} \right\} {\rm GeV\; s}^{-1} . \label{P-ion} \ee


The energy loss by pion production is given as (Mannheim and
Schlickeiser 1994, Schlickeiser 2002, p. 125 and 138)  \be -\left(
\frac {dE}{dt}\right)_{{\rm Pion},p} \sim 5.85 \times 10^{-16} \,
\left( \frac {n}{{\rm cm}^{-3}}\right) \, \left(\frac{E_p - m_p
c^2}{\rm GeV}\right)\, \Theta(E_p-E_{\rm th}) \, {\rm GeV\; s}^{-1}
. \label{P-2}\ee

\noindent {\bf Electron losses \label{ELOS}}



In the ultrarelativistic case ($E \gg mc^2$), the ionization losses
in neutral atomic matter (e.g., Schlickeiser 2002, p. 99; Ginzburg
\& Syrovatskii 1964, p. 140ff) \be  -\left( \frac{dE}{dt}
\right)_{{\rm Ion},e}  \sim 2.75 \times 10^{-17} \left[6.85 + \ln
\left(\frac{E}{mc^2} \right)\right] \left[\frac{n_{\rm H}+2 n_{{{\rm
H}}_2} }{{\rm cm}^{-3}}\right] {\rm GeV \, s}^{-1} \label{E-0}.\ee


Synchrotron losses can be computed as (e.g., Ginzburg \& Syrovatskii
1964, p. 145ff; Blumenthal \& Gould 1970) \be -\left( \frac{dE}{dt}
\right)_{{\rm Sync},e} = \frac 23 c \left( \frac{e^2}{mc^2}
\right)^2 B_{\bot}^2 \left( \frac{E}{mc^2} \right)^2 \sim 2.5 \times
10^{-6} \left( \frac{B}{{\rm Gauss}} \right)^2 \left( \frac{E}{{\rm
GeV}} \right)^2 {\rm GeV\;s}^{-1},
 \label{E-1}
\ee where $B_{\bot}$  represents the magnetic field in a direction
perpendicular to the electron velocity, and the second equality
takes into account that an isotropic distribution of pitch angles.
In this case, particles velocities are distributed according to
$p(\alpha)d\alpha= [(1/2) \sin \alpha)]  d\alpha$, with $\alpha$
the angle between the particle's velocity and $B$, varying between
0 and $\pi$. Then, as $B_{\bot}=B\sin \alpha$, the average in Eq.
(\ref{E-1}) requires the integral $\int [(1/2) \sin \alpha)] \sin
^2\alpha \, d\alpha=2/3$, in order to go from $B_\bot$ to $B$.


The losses produced by Inverse Compton emission are given by
(e.g., Blumenthal \& Gould 1970) \be \label{ic1} - \left(
\frac{dE}{dt} \right)_{{\rm IC},e}=
\int_0^\infty  d\epsilon \, \int_{{E_{\gamma}}^{\rm
min}}^{{E_{\gamma}}^{\rm max}} \, dE_\gamma \, E_\gamma c\,n_{\rm
ph}(\epsilon) \, \frac{d\sigma(\epsilon, E_\gamma, E)}{dE_\gamma}
\ee where $n_{\rm ph}(\epsilon)$ is the target photon distribution
(usually a black or a greybody),
%
%
$\epsilon$ and $E_\gamma$ are the photon energies before and after
the Compton collision, respectively, and $d\sigma(\epsilon,
E_\gamma, E)/dE_\gamma$ is the Klein-Nishina differential cross
section (Schlikeiser 2002, p. 82).
%
%
%
%
%


Additional losses are caused by the emission of bremsstrahlung
$\gamma$-ray quanta in interactions between electrons and atoms of
the medium.  The energy loss can be computed as (e.g., Schlickeiser
2002, p. 95ff; Ginzburg \& Syrovatskii 1964, p. 143, Blumenthal \&
Gould 1970): \be \label{b1} - \left( \frac{dE}{dt} \right)_{{\rm
Brem},e} = \int dE_\gamma \; E_\gamma
\left(\frac{dN}{dt\,dE_\gamma}\right), \ee where
$({dN}/{dt\,dE_\gamma})= c \sum_j n_j (d\sigma_j/dE_\gamma)$
represents the number of photons emitted with energy $E_\gamma$ by a
single electron of initial energy $E$ in a medium with $j$ different
species of corresponding densities $n_j$, and where
$(d\sigma_j/dE_\gamma)$ is the Bethe-Heitler differential cross
section.



\noindent {\bf  Leptonically-generated high energy radiation }


The bremsstrahlung emissivity can be computed from the steady CR
electron spectrum as the integral  as $ {Q_\gamma(E_\gamma)}_{{\rm
Brem}} =n\, {E_\gamma}^{-1} \int_{{E_\gamma}}^\infty   dE_e \,c\,
N_e(E_e) \, \sigma_{{\rm Brem}},$ where $\sigma_{{\rm Brem}}$ is the
bremsstrahlung cross section, equal to $3.38 \times 10^{-26}$
cm$^2$, and $n=(n_{\rm H}+2 n_{{{\rm H}}_2})$ is the ISM atomic
hydrogen density.



The inverse Compton emissivity is given by \be \label{ic2} {Q_\gamma(E_\gamma)}_{{\rm IC}} =
\int_0^\infty n_{\rm ph}(\epsilon) d\epsilon \int_{E_{\rm
min}}^{E_{\rm max}}
\frac{d\sigma(E_\gamma,\epsilon,E_e)}{dE_\gamma}\, c\, N_e(E_e) dE_e
\;
.\ee ${E_{\rm max}}$ is the maximum electron energy for which the
distribution $N_e(E_e)$ is valid. ${E_{\rm min}}$ is the minimum
electron energy needed to generate a photon of energy $E_\gamma$,
i.e. ${E_{\rm min}}=(E_\gamma/2) [1+(1+(mc^2)^2/\epsilon
E_\gamma)^{1/2}]$.\footnote{A fixed $E_{\rm max}$ implies that, for
a given resulting upscattered photon energy, there is also a minimum
energy for the photon targets in the first integral of the IC flux.
Target photons with less than this energy do not contribute to the
flux at the upscattered energy in question.}

\noindent {\bf  Synchrotron emission}

The synchrotron emissivity can be written as \ba \label{radio11}
\epsilon_{\rm Sync}(\nu) = 1.166 \times 10^{-20} \left(\frac{B}{\rm
Gauss}\right) \int dE\, N(E) {\int_0}^{\pi/2} d\alpha \,  \frac
{\nu}{\nu_c} \sin^2\alpha \int_{\nu/\nu_c}^\infty
d\xi K_{5/3}(\xi) \; \nonumber \\
\hspace{5cm} {\rm GeV\, s^{-1}\, cm^{-3}\, Hz^{-1}\, sr^{-1} }. \ea
A useful result is given by the product of $\epsilon_{\rm Sync}$ and
$V/D^2$, $f_{\rm Sync}(\nu)$. This is the synchrotron flux density
(units of Jy) expected from a region of volume $V$ located at a
distance $D$ in cases in which opacities are negligible, see below.
%
%
%
In cases where opacities are not negligible, one has to solve first
for the specific intensity considering all absorption processes,
compute the emissivity, and consider the geometry.

\noindent  {\bf  Free-free emission and absorption}

The emission  and absorbtion coefficients  for this process are
given by the following expressions (e.g., Rybicki \& Lightman 1979,
Ch. 5, Schlickeiser 2002, Ch. 6)
%
%
$ \epsilon_{\rm ff}(\nu) = 3.37 \times 10^{-36} Z^2 \left( {n_e
n_i}/{\rm cm^{-6}}\right) \left({T}/{\rm K}\right)^{-1/2}
\left({\nu}/{\rm GHz}\right)^{-0.1} e^{-h\nu/kT} {\rm
GeV\,cm^{-3}\,s^{-1}\,Hz^{-1}\,sr^{-1}} $ and $ \kappa_{\rm ff}(\nu)
= 2.665 \times 10^{-20} Z^2 \left({T}/{\rm K}\right)^{-1.35} \left(
{n_e n_i}/{\rm cm^{-6}}\right) \left({\nu}/{\rm GHz}\right)^{-2.1}
{\rm cm^{-1}}, $ respectively. Here, the plasma is described by a
temperature $T$, metallicity $Z$ and thermal electron and ion
densities $n_e$ and $n_i$, respectively. The free-free opacity is
given by the integral $ \tau_{\rm ff} \equiv \int_0^\infty dr\,
\kappa_{\rm ff} \sim 8.235\times 10^{-2} \left({T}/{\rm
K}\right)^{-1.35} \left({\nu}/{\rm GHz}\right)^{-2.1} \left( {\rm
EM}/{\rm cm^{-6} \, pc} \right), $ where EM is the emission measure,
defined as EM=$\int_0^\infty dr\, n_i n_e$. For simplicity, and in
lack of other knowledge, it is assumed that the EM is constant.
The turnover frequency $\nu_t$ (for frequencies less than $\nu_t$
the emission is optically thick) can also be given in terms of EM,
$\nu_t=0.3 [({T}/{\rm K})^{-1.35} {\rm EM}]^{1/2}$ GHz.
%

%




\noindent {\bf Radiation transport equation, and fluxes from
emissivities}

This paper analyzes the case in which emission and absorption are
uniform, co-spatial, and without further background or foreground
sources or sinks (see, e.g., Appendix A in Schlickeiser 2002). The
solution to the radiation transport equation in these situations is
$I_\nu = \frac{ \epsilon_\nu }{ \kappa_\nu } (1-e^{-\tau_\nu} ) ,
\label{RT}$ where $\epsilon_\nu$ is the emission coefficient --or
emissivity--, $\kappa_\nu$ is the absorption coefficient, and
$\tau_\nu=\kappa_\nu L$ is the opacity in the far end ($L$) of the
emission region (also referred to as the maximum opacity). In cases
in which there are more than one process involved in the emission or
in the absorption, a sum over processes must be performed.
Units are consistent with the rest of the paper, such that
$[\epsilon_\nu]=$ GeV cm$^{-3}$ s$^{-1}$ sr$^{-1}$ Hz$^{-1}$ (in the
case of $\gamma$-rays, photon emissivities are used instead,
$Q/4\pi$, with units of photons cm$^{-3}$ s$^{-1}$ sr$^{-1}$
GeV$^{-1}$), $[\kappa_\nu]=$cm$^{-1}$, and $[\tau]=1$. Additionally,
$I_\nu$ is the emergent intensity ($[I_\nu]=$GeV cm$^{-2}$ s$^{-1}$
sr$^{-1}$) after the absorption processes are considered.

Consider first the case in which opacities are negligible. To
compute the flux, given the knowledge of its emissivity under a
particular process, information on the solid angle -as seen from the
observer- ($\Omega$) and depth ($L$) along the line of sight, or
volume and distance of the region of emission is needed. For
instance, the integral flux of $\gamma$-rays, with no absorption, is
given by \be F_\gamma(E_\gamma>E)  = \int_E^\infty
{Q_\gamma(E_\gamma)} \frac{ \left[\Omega L\right]_{\rm
obs}}{4\pi}dE_\gamma
= \frac{ V}{4\pi D^2} \int_E^\infty {Q_\gamma(E_\gamma)} dE_\gamma
, \ee where $ {Q_\gamma(E_\gamma)}= {Q_\gamma(E_\gamma)}_{{\rm
Brem}} + {Q_\gamma(E_\gamma)}_{{\rm IC}} +
{Q_\gamma(E_\gamma)}_{\pi^0} $ is the total $\gamma$-ray emissivity,
and ${ \left[\Omega L\right]_{\rm obs}}/{4\pi}$ corrects for the
fraction of the emission which is in the direction of the observer.
Clearly, in this case, the differential photon flux is $
F_\gamma(E_\gamma) =[ { V}/{4\pi D^2}] {Q_\gamma(E_\gamma)}$.

\begin{figure*}[t]
\centering
\includegraphics[width=.4\textwidth]{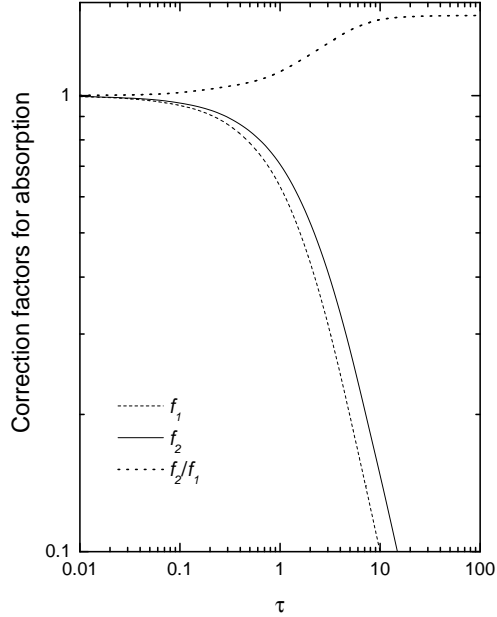}\hspace{0.5cm}
\caption{Correction factors for absorption. The $f_2/f_1$ curve
asymptotically tends to 1.5. } \label{opa1}
\end{figure*}

When there are absorption processes involved, but the geometry is
such that $I$ is not depending on the position within the emitting
region, i.e., when both emission and absorption coefficients are
uniform and the maximum value of $\tau$ is the same for all the
region\footnote{In the case of a uniform absorption coefficient this
imposes a constraint on the geometry. For example, for a
molecular disk, the linear size in the direction of the observer may
be considered the same, and thus $\tau$ is independent of any angle,
and so is $I$. }, the flux can be computed  as \be F_\nu = \frac{
\epsilon_\nu }{ \tau_\nu } (1-e^{-\tau_\nu} ) \frac{V}{D^2} \equiv
\epsilon_\nu \frac{V}{D^2} f_1. \label{FFdisk} \ee However, in the
case of an sphere, for example, even when emission and absorption
are uniform, the specific intensity is not. Because the linear size
is different at different angles $\theta^\prime$ as measured from
the center of the sphere, the opacity will also change. This change
can be represented as $\tau_\nu = \kappa_\nu \times 2R
\cos\theta^\prime = \tau_{\rm max} \cos\theta^\prime$, i.e. through
the use of the maximum opacity $\tau_{\rm max}$ affecting a photon
equatorially traversing the system. $\tau_{\rm max}$ is also a
function of the frequency, although the subindex $\nu$ is omitted
for simplicity.
The flux is \ba F &=& \int I [\cos\theta]\ d\Omega = \int \frac{
\epsilon_\nu }{ \kappa_\nu } (1-e^{{-\tau_\nu}(\theta^\prime)} )
\,2\pi\,\cos\theta \, \sin\theta \,
d\theta \nonumber \\
&=& \frac{ \epsilon_\nu }{ \kappa_\nu } 2\pi \int_0^{\theta_{\rm
max}} \left(1 - e^{\tau_{\rm max} \sqrt{1-(D/R)^2
\sin\theta}}\right) \cos\theta \sin\theta d\theta .\ea The solution
to this integral can be analytically obtained and after some algebra
the result can be written as \be F= \frac{ \epsilon_\nu }{ \tau_{\rm
max} } \frac{V}{D^2} \left[ \frac 32 + \frac 3{ \tau_{\rm max}^2 }
\left( (1+ \tau_{\rm max}) e^{- \tau_{\rm max}} -1 \right)
\right]\equiv \epsilon_\nu \frac{V}{D^2} f_2. \label{opaf2} \ee Note
that when $ \tau_{\rm max} \ll 1$ the previous result reduces to the
case of no absorption, $f_2=1$. Figure \ref{opa1} shows the behavior
of the correction factors for absorption that appear in the
different contexts analyzed in this paper, $f_1$ and $f_2$.

\noindent {\bf Dust emission}

We assume that the dust photon emissivity, which dominates the
luminosity at micron-frequencies, is given by $q_d=q_0
\epsilon^\sigma B(\epsilon,T)$, where $\sigma \sim 1-2$ is the
emissivity index, $B(\epsilon,T)$ is the Planck function of
temperature $T$, and $\epsilon$ is the photon energy (see, e.g.,
Rice et al. 1988; Goldshmidt and Rephaeli 1995; Kr\"ugel 2003,
p.245).  Units correspond to $[q_d]=$ photons s$^{-1}$ cm$^{-2}$.
Then, the flux produced by dust can be computed as $F=2\pi
\int_0^{\pi/2} q_d \cos \theta \sin \theta d\theta d\epsilon=\pi
\int  q_d d\epsilon$ and normalized to $[L/4\pi R^2]$, with $L$ and
$R$ being the IR luminosity and radius of the emitting region,
respectively; i.e., normalized to the power per unit area through
the surface of the emitting region. This fixes the dimensional
constant. Units are such that $[q_0]=$ GeV$^{-1-\sigma}$ s$^{-1}$,
and [$B(\epsilon,T)]=$ GeV cm$^{-2}$.

The flux density of dust emission at the surface of the emitting
region is obtained from the definition $F \equiv \int f_{\rm
dust}(\nu) d\nu$, where units are, for consistency, $[f_{\rm
dust}]=$ s$^{-1}$ cm$^{-2}$ Hz$^{-1}$ GeV. The IR photon number
density per unit energy, $n(\epsilon)$, can be obtained by equating
the particle flux outgoing the emission region, $\pi R^2 c
n(\epsilon) \epsilon d\epsilon$, with the expression of the same
quantity that make use of the emissivity law, $4\pi R^2 \pi
q(\epsilon) d\epsilon$.

\end{document}